\documentclass[journal]{IEEEtran}

\usepackage[T1]{fontenc}
\usepackage[utf8]{inputenc}
\usepackage{amsmath,amssymb,bm}
\usepackage{graphicx}
\usepackage{booktabs}
\usepackage{array}
\usepackage{siunitx}
\usepackage[numbers,sort&compress]{natbib}
\usepackage{url}
\usepackage{xcolor}
\usepackage[colorlinks=true,allcolors=blue!55!black]{hyperref}

\newcommand{\snot}{\sigma^{0}}
\newcommand{\mss}{\mathrm{mss}}
\newcommand{\msseff}{\mathrm{mss}_{\mathrm{eff}}}
\newcommand{\Uten}{U_{10}}
\newcommand{\Hs}{H_s}
\newcommand{\dB}{\,\mathrm{dB}}
\newcommand{\qz}{q_z}
\newcommand{\qh}{q_H}
\newcommand{\rstar}{r^{*}}
\newcommand{\Dtwo}{D_2}

\graphicspath{{figures/}}

\begin{document}
\bstctlcite{IEEEexample:BSTcontrol}

\title{A Dual-Frequency-Constrained Structure-Function\\
Kirchhoff Model for Near-Nadir Ocean Backscatter\\
---Part~I: Theoretical Framework}

\author{Alexey~S.~Mironov,
        Fr\'ed\'eric~Nouguier,
        and~Bertrand~Chapron%
\thanks{A.~S.~Mironov is with eOdyn, Plouzan\'e, France
        (e-mail: alexey.mironov@eodyn.com).}%
\thanks{F.~Nouguier and B.~Chapron are with the Laboratoire d'Oc\'eanographie
        Physique et Spatiale (LOPS), IFREMER, Plouzan\'e, France
        (e-mail: frederic.nouguier@ifremer.fr; bchapron@ifremer.fr).}%
\thanks{This work has been submitted to the IEEE for possible publication. Copyright may be transferred without notice, after which this version may no longer be accessible.}}

\markboth{Preprint submitted to IEEE Transactions on Geoscience and Remote Sensing}%
{Mironov \emph{et al.}: Structure-Function Kirchhoff Model for Near-Nadir Backscatter --- Part I}

\maketitle

\begin{abstract}
Most operational near-nadir geophysical model functions (GMFs) are empirical and fitted one
radar band at a time, and do not explain the difference between Ku and Ka band.
This paper derives a backscatter model in which one local description of the sea surface serves both
bands. The near-nadir Kirchhoff integral is written in terms of the elevation difference between two
surface points separated by a horizontal lag. The surface then enters through the structure functions of
that difference, and the scattering kernel selects the lags each band samples, about $2.5$--$1.4$\,cm at
Ku and $1.0$--$0.5$\,cm at Ka band. Gaussian, isotropic geometric optics is recovered
for a quadratic structure function, without a spectral cutoff. With a logarithmic second-order structure
function, one wind-dependent log-slope sets both the incidence shape and the Ku--Ka shape difference, and
band transfer reduces to evaluating one curve at a second lag. Same-footprint dual-frequency radar
data show that one band alone does not determine this transfer. The log-slope is the level
of the short-wave saturation range and grows about as the square of the wind speed; a physical short-wave
spectrum reproduces this growth but overestimates the level. Dual-frequency radars thus measure the
centimetre-scale roughness level, although they do not separate a scale-dependent structure function from
non-Gaussian surface statistics. A companion paper delivers and evaluates the resulting GMF, KuKa-Mod.
\end{abstract}


\section*{Nomenclature}
\begin{IEEEdescription}[\IEEEsetlabelwidth{$D_{22},\delta_a,\delta_b$}]
\item[$\Dtwo(r)$] second structure function $\langle\Delta\eta^2\rangle$ of the elevation increment $\Delta\eta=\eta(\bm x+\bm r)-\eta(\bm x)$ at lag $r$; $D_3$, $D_4$ are the higher cumulants.
\item[$s^2(r)$] lag-dependent slope variance $2\Dtwo(r)/r^2$.
\item[$a(\Uten)$] additive coefficient of the logarithmic form \eqref{eq:marginallog}.
\item[$b(\Uten)$] small-lag log-slope of $s^2(r)$, the short-wave saturation level.
\item[$b_\Delta$] shape-difference log-slope: the $b$ that reproduces the matched Ku--Ka shape difference alone.
\item[$r_{\rm ref}$] reference lag of $s^2(r)$, fixed at $1$\,m (convention).
\item[$\qz,\qh$] vertical and horizontal components of the electromagnetic scattering vector, $2k_r\cos\theta$, $2k_r\sin\theta$.
\item[$\rstar$] kernel e-folding lag, $\Dtwo(\rstar)=2/q_z^2$.
\item[mss] total mean-square slope of the surface.
\item[$\msseff$] radar-effective (filtered) mean-square slope, $s^2(\rstar)$.
\item[$E(k)$, $B(k)$] omnidirectional elevation spectrum and saturation spectrum $B=k^3E$.
\item[msc] mean-square curvature $\int k^4E\,dk$, in $\mathrm{m^{-2}}$; distinct from $b$.
\item[$R$] variance-mixture kurtosis parameter (excess kurtosis $3R$).
\item[$\lambda_3$] amplitude of the odd closure $D_3=\lambda_3D_{20}^{\,\nu}\cos\psi$ (Sec.~\ref{sec:d3}); $D_3(0)=0$.
\item[$\lambda_3^{0}$] amplitude of the regularized $D_3$ closure \eqref{eq:d3reg}; the data fix $\lambda_3^{0}\sqrt{D_c}$.
\item[$\psi$] azimuth of the lag $\bm r$ relative to the wind.
\item[$\lambda_4$] Gram--Charlier excess kurtosis of the slope; $\lambda_4=3R$ for the scale mixture.
\item[$\Phi$] wind-relative azimuth, $\Phi=0$ looking downwind.
\item[$A_1,A_2$] linear multiplicative wind-relative azimuth harmonics of $\snot$ (Eq.~\eqref{eq:full}).
\item[$a_1,a_2$] the same harmonics in dB, $a_n\simeq(10/\ln10)A_n$.
\item[$R_{\rm eff}^2$] nadir level factor $R_F^2(f,T,S)\,C_f$.
\item[$D_{20}$] isotropic (azimuth-averaged) part of $\Dtwo$.
\item[$D_{22}$, $\delta_a,\delta_b$] $\cos2\psi$ anisotropy of $\Dtwo$ and its two coefficients (Sec.~\ref{sec:complete}).
\item[$\sigma^2_{\rm long}$] slope variance of the long waves, $\int_{k<3\,\mathrm{rad\,m^{-1}}}k^2E\,dk$.
\item[$\Hs$] significant wave height.
\item[$\Delta\sigma^2_{\rm long}$] long-wave slope variance in excess of its value at the reference sea state, $\Hs\in[1.5,3.5]$\,m.
\item[$|M|$, $m_s$] modulation transfer amplitude of the radar-band mss per unit long-wave slope, and its straining coefficient $d\ln\mss/d\delta$.
\item[$\mss_w$] effective slope variance $\msseff$ at the reference sea state.
\item[$H$, $D_{\rm F}$] local Hurst exponent of $\Dtwo\propto r^{2H}$ and fractal dimension $D_{\rm F}=3-H$.
\item[$\sigma_U$, $\ell$] standard deviation of the $10$-m wind within a footprint of size $\ell$ (subgrid term).
\end{IEEEdescription}

\section{Introduction}
\label{sec:intro}

The radar echo returned by the ocean near vertical incidence reads how the wind
roughens the surface, through slope statistics sampled over centimetre-scale lags. This echo has been recorded from space for more than four
decades. Altimeters convert their nadir cross-section into wind speed
\citep{chelton1986,gourrion2002}. The precipitation radars of the Tropical
Rainfall Measuring Mission (TRMM) and the Global Precipitation Measurement (GPM)
mission use the same ocean return as the reference for rain path attenuation, and
also to measure wind \citep{meneghini2000,freilich2003,hou2014}. The Surface Waves Investigation and Monitoring (SWIM) wave
scatterometer on the China--France Oceanography Satellite (CFOSAT) derives
directional wave spectra from its modulation \citep{hauser2017,hauser2021}.

Each of these retrievals passes through a geophysical model function (GMF) that
relates the normalized cross-section $\snot$ to wind, incidence, azimuth and sea
state, and each GMF is tied to one radar band. Most of the instruments above
operate at Ku band. Ka band is carried by the AltiKa altimeter and by the Surface Water and
Ocean Topography (SWOT) swath altimeter, while the GPM radar and the planned
Copernicus polar ice and snow topography altimeter (CRISTAL) combine
both bands \citep{verron2015,morrow2019,kern2020}. A dual-frequency radar thus
sees the same sea surface at two wavelengths, and this information on the short
waves remains largely unused. A GMF that describes both bands with one sea
surface is therefore needed, both to inter-calibrate the missions and to read the
band difference as information on the short waves. Ku/C altimeters have flown since TOPEX,
and a Kirchhoff analysis of that pair \citep{chapron1995} gave the initial form of the
spectrum of \citet{elfouhaily1997}.

Below about $12^\circ$ incidence the return is dominated by specular reflection
from surface facets oriented perpendicular to the radar beam. As the wind roughens
the surface, fewer facets face the radar, so near nadir $\snot$ decreases with wind and its
information lies in the slope statistics. Most operational near-nadir GMFs
nevertheless describe this dependence empirically. They fit $\snot$ to collocated
model or buoy winds, one band and one instrument at a time
\citep{chelton1986,gourrion2002,ren2016,hossan2021}, without an explicit scattering
mechanism. What such a fit learns about the sea surface in one band therefore
cannot be carried to another band. A common physically based alternative is filtered-slope geometric
optics (GO), with a slope variance obtained by integrating the wave spectrum up to
a cutoff tied to the radar wavelength
\citep{brown1978,jackson1992,vandemark2004,nouguier2016}. In that approach the
cutoff is a convention, and the non-Gaussian peakedness of the slope distribution
enters as an empirical correction. A second band can then be reached only through
an assumed short-wave spectrum. This limitation becomes decisive with matched
observations, in which GPM sees the same footprint at Ku and Ka: one sea surface
should then explain both bands at once.

The present paper builds a GMF whose retrieved quantities describe the sea surface
and therefore transfer between bands. Written through the elevation difference
between two points at a horizontal lag, the near-nadir Kirchhoff integral contains
Gaussian, isotropic geometric optics as its quadratic, zero-higher-cumulant limit, and a lag selected by the
scattering kernel replaces the spectral cutoff. Ku and Ka band thus read one structure
function at two lags, and their difference measures the short waves.

This forward formulation builds on classical work. The Kirchhoff equation and its
expansion in elevation statistics are standard \citep{barrick1968,brown1978,mouche2007}.
Its fourth-order form, which includes the mean-square curvature, and the
corresponding one-band inversion were derived by \citet{nouguier2009phd} and
published as the GO4 model by \citet{boisot2015}. Within that framework,
\citet{nouguier2016} separated a band-invariant kurtosis from a band-dependent
curvature. Related work developed non-Gaussian closures
\citep{tatarskii1999b,tatarskii2003} and related dual-frequency altimetry to the
fractal dimension of the sea \citep{agnon1991}. Empirical Ku and Ka GMFs also
exist, but each band is fitted separately \citep{hossan2021}. What the present
paper adds is the inverse step: one common local structure function retrieved from
matched Ku/Ka observations.

This inverse step leads to three results. The scattering kernel selects
centimetre-scale lags, at which the filtered mean-square slope becomes the slope
variance. One log-slope $b(\Uten)$ sets both the incidence shape and the Ku--Ka shape
difference, which a single band does not fix. A physical short-wave spectrum gives the
wind exponent of $b$, close to $\Uten^{2}$, but not its level. Two frequencies do not
separate a scale-dependent structure function from a non-Gaussian closure, so an
independently fitted fourth cumulant is optional. Numerical illustrations use the
parameters retrieved jointly from GPM Ku and Ka in Part~II \citep{part2}, called the
joint parameters below.

\section{The Structure-Function Kirchhoff Formulation}
\label{sec:formulation}

The starting point is the Kirchhoff, or tangent-plane, approximation, which requires
surface curvature radii large compared with the radar wavelength but is not
restricted to near-nadir incidence. The scattering amplitude over an illuminated
area $A$ is the surface integral
\begin{equation}
  S(\bm k,\bm k_0)=\frac{\mathcal K}{\qz}\,\frac{1}{(2\pi)^2}
    \int_A d\bm r\;e^{-i\bm\qh\cdot\bm r}\,e^{\,i\qz\eta(\bm r)},
  \label{eq:amplitude}
\end{equation}
with $\mathcal K$ the polarization- and geometry-dependent Kirchhoff kernel
\citep{elfouhaily2004,nouguier2009phd}. For a statistically homogeneous surface
$\eta(\bm x)$, the ensemble-averaged square of \eqref{eq:amplitude} gives, as
$A\to\infty$, the backscatter cross-section \citep{barrick1968,brown1978}
\begin{equation}
  \snot(\theta)=\frac{k_r^2 R_{\rm eff}^2}{\pi\cos^2\theta}
    \int d\bm r\;e^{-i\bm\qh\cdot\bm r}\,
    \big\langle e^{\,i\qz\Delta\eta(\bm r)}\big\rangle.
  \label{eq:master}
\end{equation}
Here $\theta$ is the incidence angle and $k_r$ the radar wavenumber, and
$\qz=2k_r\cos\theta$ and $\qh=2k_r\sin\theta$ are the vertical and horizontal
components of the electromagnetic scattering vector. The effective nadir
reflectivity $R_{\rm eff}^2$ combines the Fresnel power reflectivity with a
calibration, path and unresolved-roughness factor (Section~\ref{sec:bandentry}). The elevation increment
over the horizontal lag $\bm r$ is $\Delta\eta(\bm r)=\eta(\bm x+\bm r)-\eta(\bm x)$,
and the coherent component has been subtracted.

The surface thus enters \eqref{eq:master} only through the two-point
characteristic function $\langle e^{i\qz\Delta\eta}\rangle$, which reduces the
joint statistics of the elevation at two points to the distribution of their
increment. Expanding that function in cumulants makes its physical content
explicit:
\begin{equation}
  \begin{aligned}
  \big\langle e^{\,i\qz\Delta\eta(\bm r)}\big\rangle
  = \exp\big(&-\tfrac12\qz^2\Dtwo(\bm r)
      -\tfrac{i}{6}\qz^3 D_3(\bm r)\\
      &+\tfrac{1}{24}\qz^4 D_4(\bm r)+\dots\big).
  \end{aligned}
  \label{eq:cumulants}
\end{equation}
Here $D_n(\bm r)=\langle\Delta\eta^n\rangle_c$, the subscript $c$ denoting a
cumulant, are the generalized structure functions of the surface. Each has a
distinct role. The second, $\Dtwo$, is the ordinary structure function, in general
anisotropic. The third, $D_3$, changes sign under lag reversal $\bm r\to-\bm r$ and
therefore contributes an imaginary phase. Off nadir, this phase produces the odd
($\cos\Phi$) azimuthal modulation of $\snot$ assembled in \eqref{eq:full}. The
fourth, $D_4$, sets the excess kurtosis of the increment, $D_4/\Dtwo^2$, which in the
small-lag geometric-optics limit becomes the peakedness of the slope distribution.
The expansion also contains a known result. The product of the quadratic term and
the cubic phase in \eqref{eq:cumulants} is exactly the third-order skewness
correction to the Kirchhoff characteristic function obtained by
\citet{mouche2007}, here written in elevation-increment cumulants instead of slope
moments.

Within the scalar Kirchhoff approximation, the roughness content of the near-nadir
GMF is therefore the set of structure functions $\{D_n(\bm r)\}$, evaluated at the
lags that the scattering integral weights. We build the model one cumulant at a
time and give each a physical reading, in the order $\Dtwo$
(Section~\ref{sec:d2}), $D_4$ (Section~\ref{sec:d4}) and $D_3$
(Section~\ref{sec:d3}). For an isotropic surface the angular integral in
\eqref{eq:master} reduces to a Hankel transform,
\begin{equation}
  \snot(\theta)=\frac{2k_r^2 R_{\rm eff}^2}{\cos^2\theta}
    \int_0^\infty r\,J_0(\qh r)\,
    \hat\chi\!\big(\qz;\,r\big)\,dr,
  \label{eq:hankel}
\end{equation}
where $J_0$ is the Bessel function of order zero and $\hat\chi(\qz;r)$ the
isotropic, even-cumulant part of the characteristic function \eqref{eq:cumulants}.
The directional terms are left out of \eqref{eq:hankel} on purpose. The odd $D_3$
has no isotropic part, and the $\cos2\psi$ anisotropy of $\Dtwo$, with $\psi$ the
lag azimuth relative to the wind, averages out over azimuth. Both re-enter through
Bessel kernels of order one and two as the azimuth harmonics assembled in
Section~\ref{sec:complete}.

One more term is absent from \eqref{eq:hankel}, the coherent return
of the whole surface, $e^{-\qz^2\sigma_\eta^2}$ for Gaussian elevations of variance
$\sigma_\eta^2$. It is negligible, since $\qz^2\sigma_\eta^2\gtrsim10^3$ at Ku and Ka
for the wind seas considered here. Roughness unresolved within a facet is another matter. Independent unresolved roughness enters $\Dtwo$
as a nugget, $\Dtwo(r)=\Dtwo^{\rm res}(r)+c_0$ for $r>0$ with $\Dtwo(0)=0$: the
resolved part stays smooth near the origin, and the jump at $r=0$ stands for a narrow
covariance contribution. The nugget multiplies the return by $\exp(-\qz^2c_0/2)$, a
reduction of the effective power reflectivity whose nadir value is absorbed in
$R_{\rm eff}^2$ (Section~\ref{sec:bandentry}) and whose angular dependence is weak near
nadir. The non-saturating $\Dtwo$ adopted below drives $\hat\chi\to0$ at large lag.
Equation~\eqref{eq:hankel} is the working form of the model, and everything that
follows specifies its integrand.

\section{The Geometric-Optics Limit and the Kernel Lag}
\label{sec:golimit}

Geometric optics does not require Gaussian statistics. Where the surface is smooth
over the kernel lags, $\Delta\eta\simeq\bm r\cdot\nabla\eta$, and \eqref{eq:master}
reduces to the specular-point form
$\snot_{\rm GO}=\pi R_{\rm eff}^2\,p(\bm q_h/\qz)/\cos^4\theta$ for any slope
density $p$, non-Gaussian or anisotropic. The classical GO GMF is its Gaussian,
isotropic case. In the working form \eqref{eq:hankel}, $\snot\propto\int_0^\infty
r\,J_0(\qh r)\,\hat\chi(\qz;r)\,dr$, an exactly quadratic and isotropic
$\Dtwo(r)=\tfrac12\mss\,r^2$ with $D_n=0$ for $n>2$ makes the Hankel integral elementary
and returns, in closed form,
\begin{equation}
  \snot_{\rm GO}(\theta)=\frac{R_{\rm eff}^2}{\mss\cos^4\theta}
    \exp\!\Big(-\frac{\tan^2\theta}{\mss}\Big).
  \label{eq:go}
\end{equation}
Gaussian GO is thus recovered when $\Dtwo(r)$ remains quadratic and isotropic over
the lags that dominate the scattering integral and the higher cumulants vanish. With the
higher cumulants still zero, a quadratic anisotropic $\Dtwo$ gives the elliptical Gaussian slope law,
and non-zero slope cumulants give the Gram--Charlier forms of \citet{cox1954,coxmunk1954}.

Here the anisotropy is a perturbation: the $\cos2\psi$ part of $\Dtwo$ and the odd
$D_3$ are linearized into the azimuth harmonics of Section~\ref{sec:complete}, which
reproduces the elliptical GO law to first order and leaves an
$O[(\Delta\mss/\mss)^2]$ effect on the azimuth-averaged profile, negligible at the
measured few-percent anisotropy.

The structure-function view also explains why GO falls short near nadir. Filtered-slope
GO needs a spectral cutoff to define the radar slope variance, and a prescribed slope
density carries peakedness but does not fix the transfer between bands. The kernel
lag replaces the cutoff even for a quadratic $\Dtwo$. A nonquadratic $\Dtwo$
supplies the band transfer and a finite-wavelength part of the peakedness
(Section~\ref{sec:disc-go}). Higher cumulants are a second, statistical source.

\begin{figure*}[t]
  \centering
  \includegraphics[width=\textwidth]{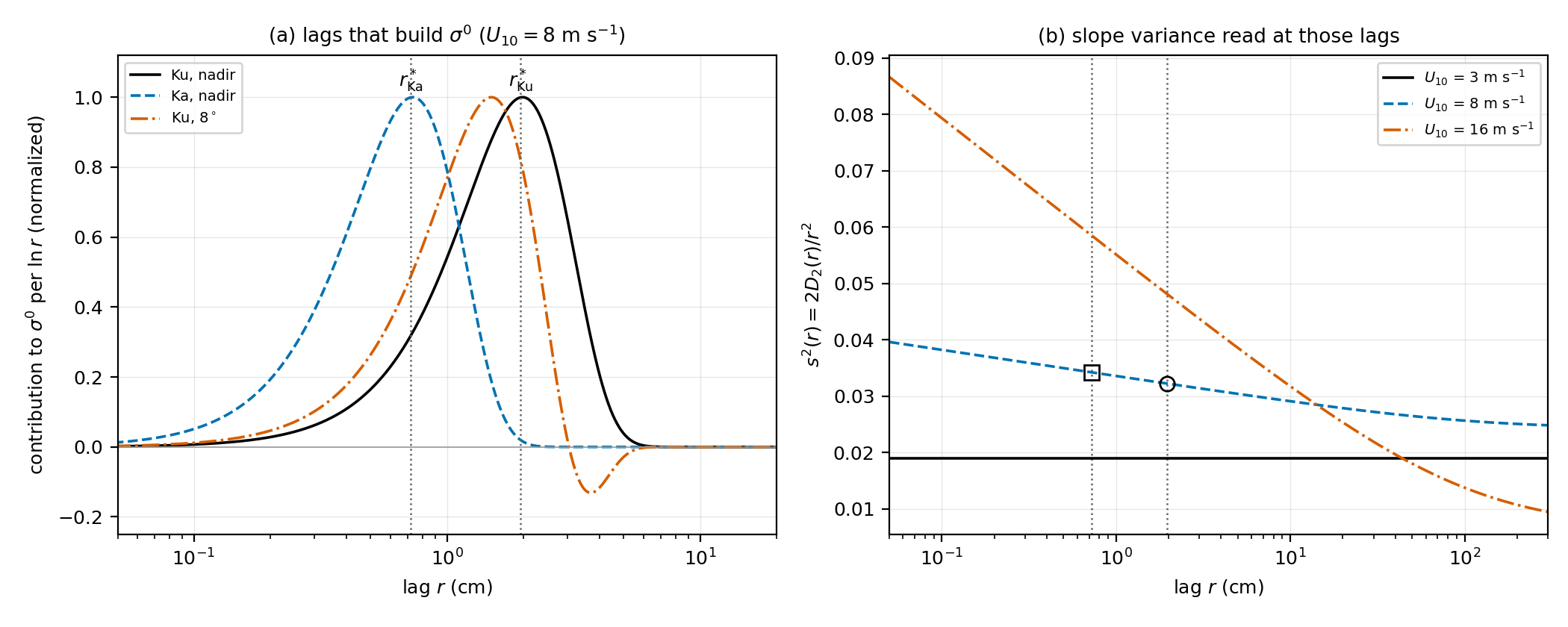}
  \caption{Mechanics of the structure-function Kirchhoff model (joint parameters,
  $\Uten=8\,\mathrm{m\,s^{-1}}$). (a)~Contribution of each lag to $\snot$, per unit
  $\ln r$ and normalized to its peak. At nadir it peaks near the kernel e-folding lag
  $\rstar$ (dotted), $1.96$\,cm for Ku and $0.72$\,cm for Ka, and falls to half at
  about $0.5\,\rstar$ and $1.7\,\rstar$. Off nadir (Ku, $8^\circ$) the Bessel factor
  $J_0$ shifts the peak to smaller lags and adds a negative lobe. The kernel itself
  limits the lags, so no spectral cutoff is imposed. (b)~The effective slope
  variance $s^2(r)=2\Dtwo(r)/r^2$ for three winds. The markers are the values read
  at $\rstar_{\rm Ku}$ (circle) and $\rstar_{\rm Ka}$ (square), the radar-effective
  slope variances of \eqref{eq:rstar}. Band transfer slides the marker along one
  curve. At $3\,\mathrm{m\,s^{-1}}$ the log-slope $b$ sits on its lower bound, zero, and $s^2$ is flat.}
  \label{fig:mechanism}
\end{figure*}

Which lags dominate the integral is decided by the kernel itself. The Gaussian
factor $e^{-\qz^2\Dtwo(r)/2}$ in the integrand falls to $e^{-1}$ at a lag $\rstar$
defined by
\begin{equation}
  \Dtwo(\rstar)=\frac{2}{\qz^2}
  \qquad\Longrightarrow\qquad
  \msseff\equiv s^2(\rstar)=\frac{2\Dtwo(\rstar)}{{\rstar}^2}.
  \label{eq:rstar}
\end{equation}
We call $\rstar$ the kernel e-folding lag. It is a diagnostic scale rather than a
unique ``scale measured by the radar'', because the kernel has a finite width. Per
unit $\ln r$, the contribution of each lag to $\snot$ is
$K(r)\propto r^2J_0(\qh r)\,e^{-\qz^2\Dtwo(r)/2}$. At nadir it peaks near $\rstar$ and
falls to half its peak at about $0.5\,\rstar$ and $1.7\,\rstar$
(Fig.~\ref{fig:mechanism}a). Off nadir the Bessel factor changes sign, so the
kernel acquires a negative lobe at larger lags. Its $|K|$-weighted centre
nevertheless stays within about $30\%$ of $\rstar$, so the integral averages
$\Dtwo$ over a band of lags centred near $\rstar$. For Ku band ($\lambda=2.2$\,cm)
$\rstar$ lies between $2.5$ and $1.4$\,cm over $3$--$20\,\mathrm{m\,s^{-1}}$, and for
Ka band ($\lambda=0.84$\,cm) between $1.0$ and $0.5$\,cm.

The kernel lag thus replaces the diffraction cutoff $k_d=k_r/3$ of
\citet{jackson1992}, and band transfer becomes a sliding of the kernel along one
$\Dtwo(r)$ curve (Fig.~\ref{fig:mechanism}b).

As a numerical check, \eqref{eq:hankel} with a quadratic $\Dtwo$ reproduces
\eqref{eq:go} over $0$--$18^\circ$ to $10^{-10}\dB$.

\section{The Two Channels of Band Dependence}
\label{sec:bandentry}

The kernel lag shows how the radar band enters the model, and two channels of
band dependence follow from it, one shaping the cross-section and one setting its
level. The radar wavelength enters \eqref{eq:hankel} through $k_r$ in three
places: the prefactor $k_r^2$, the vertical wavenumber $\qz=2k_r\cos\theta$, and
the horizontal wavenumber $q_H=2k_r\sin\theta$ in the Bessel factor $J_0(q_H r)$.

The first channel is the surface roughness. At fixed incidence
$q_H/\qz=\tan\theta$, so a change of radar wavelength mainly rescales the lags at
which the surface is sampled. In the dimensionless variable $\rho=\qz r$, the
Bessel factor becomes $J_0(\rho\tan\theta)$ and is the same for all bands, while
the surface is sampled at the lag $r=\rho/\qz$. The wavenumber $\qz$ thus sets the
e-folding lag $\rstar$ through \eqref{eq:rstar}, and hence the radar-effective
slope variance $\msseff=s^2(\rstar)$. A shorter wavelength moves $\rstar$ to
smaller lags and usually samples a larger effective slope variance, but it does so
by moving along the same local $\Dtwo(r)$. No per-band roughness parameter is
therefore added within the chosen $\Dtwo$ family.

\begin{figure*}[t]
  \centering
  \includegraphics[width=\textwidth]{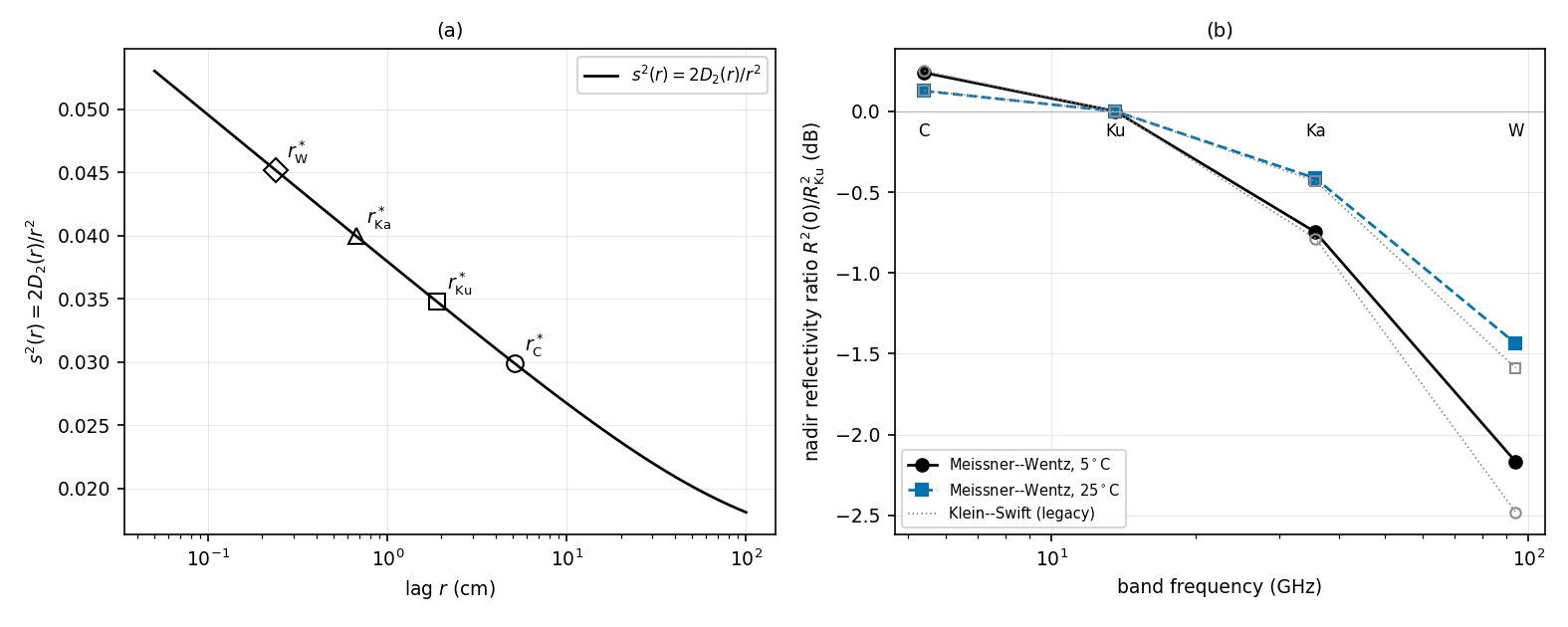}
  \caption{The two channels of band dependence, illustrated with single-band
  $\Dtwo$ parameters at $\Uten=8\,\mathrm{m\,s^{-1}}$ and the illustrative band frequencies
  $5.4$ (C), $13.6$ (Ku), $35.75$ (Ka) and $94$\,GHz (W). (a)~Roughness channel.
  Every band reads the same effective slope variance $s^2(r)=2\Dtwo(r)/r^2$ at its
  own kernel e-folding lag $\rstar$, marked at $5.1$, $1.9$, $0.67$ and
  $0.24$\,cm for C, Ku, Ka and W. C, Ku and Ka sample centimetre lags, whereas W
  approaches the millimetre range. (b)~Reflectivity channel. The nadir Fresnel
  reflectivity relative to Ku is computed from the Meissner--Wentz seawater
  dielectric for two sea-surface temperatures, with the Klein--Swift model dotted
  for comparison (open circle, $5^\circ$C; open square,
  $25^\circ$C). The Fresnel contribution to the cross-band level is thus predicted. The two
  dielectric models agree on the ratio to $0.01\dB$ (C/Ku), $0.04\dB$ (Ka/Ku) and
  $0.15$--$0.3\dB$ (W/Ku).}
  \label{fig:bandloci}
\end{figure*}

The second channel is the level, set by the nadir level factor
$R_{\rm eff}^2=R_F^2(f,T,S)\,C_f$ in the prefactor of \eqref{eq:master}. This form
separates two factors. The Fresnel power reflectivity $R_F^2$ depends on the radar
frequency $f$, the sea-surface temperature $T$ and the salinity $S$ through the
seawater dielectric. The factor $C_f$ collects the instrument calibration, the residual atmospheric
attenuation at Ka, and the loss from unresolved facet roughness (the nugget $c_0$). The dielectric model
predicts the Fresnel ratio $R_F^2(f_2)/R_F^2(f_1)$ and its
temperature dependence. It is evaluated with the double-Debye model of
\citet{meissner2004}, accurate to about $90$\,GHz. The Ka/Ku nadir reflectivity
ratio then runs from $-0.75\dB$ at $5^\circ$C to $-0.42\dB$ at $25^\circ$C
(Fig.~\ref{fig:bandloci}b), and the Klein--Swift model \citep{klein1977} agrees on it to
within $0.04\dB$. The level is thus partly computable: the dielectric predicts
$R_F^2$, and only $C_f$ is calibrated.

Together the two channels make band transfer an interpolation between the two
kernel lags. A single band measures $\Dtwo(r)$ only near its own $\rstar$, so a
model fitted in one band must extrapolate $\Dtwo$ to lags it never sampled, least
reliably at high wind, where $\Dtwo$ departs most from a single scaling law with
lag. Two bands measure $\Dtwo$ at both $\rstar_{\rm Ku}$ and $\rstar_{\rm Ka}$. In
this sense the framework is \emph{dual-frequency-constrained}:
within the assumed two-parameter family, two frequencies are the minimal set that
fixes the local scale dependence of $\Dtwo$ between the two kernel lags.
Transfer to a band outside that interval is again extrapolation. Predicting C or W
band carries the fitted logarithmic form beyond the lags bracketed by the Ku and Ka
kernels, unless measurements at that band are added to the fit. How far $\rstar$
may move before the tangent-plane description itself fails is discussed in
Section~\ref{sec:disc-validity}. Figure~\ref{fig:bandloci} shows both channels at
once: C, Ku, Ka and W band sample one effective slope-variance curve $s^2(r)$ at
their own kernel lags, and each has its own dielectric reflectivity.

\section{The Second Cumulant $\Dtwo$ and the Band Transfer}
\label{sec:d2}

Band transfer therefore depends on how $\Dtwo$ varies between the two kernel lags.
That variation follows from the shape of the short-gravity spectrum. Let $E(k)$ be
the one-dimensional omnidirectional elevation spectrum, normalized so that
$\int_0^\infty E\,dk=\langle\eta^2\rangle$ and $\int_0^\infty k^2E\,dk=\mss$. The
isotropic structure function is then $\Dtwo(r)=2\int_0^\infty[1-J_0(kr)]E(k)\,dk$.
Splitting this integral at $k\sim1/r$ separates two contributions. Long waves, with
$kr\ll1$, contribute their slope variance, since $1-J_0\simeq(kr)^2/4$. Short waves,
with $kr\gg1$, contribute only their bounded elevation variance, since $1-J_0\to1$.
In terms of the dimensionless saturation spectrum $B(k)=k^3E(k)$
\citep{phillips1985}, the split reads
\begin{equation}
  \Dtwo(r)\simeq\frac{r^2}{2}\int_{k_0}^{1/r}B(k)\,d\ln k
    +2\int_{1/r}^\infty\frac{B(k)}{k^2}\,d\ln k,
  \label{eq:split}
\end{equation}
where $k_0$ is the lower end of the wave spectrum.

In a saturation range, where $B\simeq B_0$ is roughly constant, the first integral
grows as $\ln(1/r)$ over the radar-sensitive interval. The second is smaller,
though only by a logarithmic factor at centimetre lags, so the logarithmic form
models the leading piece. The surface is then \emph{marginally fractal}. A smooth
surface, with $\Dtwo\propto r^2$ at small lags, has a finite slope variance. A
fractal surface, with $\Dtwo\propto r^{2H}$ and a Hurst exponent $H<1$, has a slope
variance that diverges as a power of $1/r$. In a saturation range it diverges only
logarithmically, because each factor of ten in wavenumber adds the same mean-square
slope, and the structure function follows the borderline logarithmic form
\begin{equation}
  \Dtwo(r)=\frac{r^2}{2}\,s^2(r),\qquad
  s^2(r)=a+b\,\ln\!\Big(1+\frac{r_{\rm ref}}{r}\Big),
  \label{eq:marginallog}
\end{equation}
in which $b$ is the dimensionless saturation level $B_0$ and the constants from
the $J_0$ expansion and the $\ln(1+\cdot)$ regularization are collected in the
coefficient $a$. Over the centimetre lags sampled here ($r\ll r_{\rm ref}$), the
local log-slope $-\partial s^2/\partial\ln r=b\,r_{\rm ref}/(r+r_{\rm ref})$ equals
$b$ to within a few percent, and $s^2(r_{\rm ref})=a+b\ln2$.
Equation~\eqref{eq:marginallog}, the logarithmic structure-function form, is the
relation on which the rest of the model is built.

How far this surface lies from smoothness can be measured by an effective fractal
dimension. With a scale-dependent Hurst exponent
$H(r)=\tfrac12\,d\ln\Dtwo/d\ln r$, the logarithmic form gives
$H(r)=1-\tfrac12\,b\,r_{\rm ref}/[(r+r_{\rm ref})\,s^2(r)]$. The fractal dimension
$D_{\rm F}=3-H$, read at the kernel lag, is then
\begin{equation}
  D_{\rm F}=2+\frac{1}{2}\,\frac{b\,r_{\rm ref}}{(\rstar+r_{\rm ref})\,s^2(\rstar)}
  \;\simeq\;2+\frac{b}{2\,\msseff},
  \label{eq:fractaldim}
\end{equation}
the last equality holding for $\rstar\ll r_{\rm ref}$. Thus $b\to0$ returns the
smooth limit $D_{\rm F}=2$, at the $E(k)\!\sim\!k^{-3}$ borderline between smooth and
fractal scaling \citep{glazman1986,agnon1991,warnick1998}. With the joint parameters,
$D_{\rm F}$ at the Ku kernel lag rises from $2.01$ at $4$ to $2.10$ at $16\,\mathrm{m\,s^{-1}}$. It
enters the classical range $2.05$--$2.10$ inferred from measured spectral slopes
\citep{banner1989,agnon1991} from about $10\,\mathrm{m\,s^{-1}}$.
An independently fitted non-Gaussian closure is a separate extension
(Section~\ref{sec:d4}); variance mixing produces a $D_4$ even without it.

Inserting \eqref{eq:marginallog} into the e-folding condition \eqref{eq:rstar}
yields a closed band-transfer law. Write $m\equiv\msseff=s^2(\rstar)$. Because
$\rstar\propto\lambda/\sqrt{m}$ to leading order,
\begin{equation}
  m(\lambda)\simeq\text{const}-b\ln(\lambda/\lambda_0)+O(\ln m),
  \label{eq:bandlaw}
\end{equation}
where $\lambda_0$ is a reference wavelength. The radar-effective slope variance
thus increases logarithmically toward shorter radar wavelengths, with a slope set
by the same $b$ that shapes the $\snot(\theta)$ profile. This leading-order law
holds for $b\ll m$ and $\rstar\ll r_{\rm ref}$. Without these approximations the
e-folding condition is implicit in $m$, and differentiating \eqref{eq:marginallog}
at fixed incidence gives
\begin{equation}
  \frac{d\,m}{d\ln\lambda}=-\frac{b\,\varphi(\rstar)}{1-b\,\varphi(\rstar)/(2m)},
  \qquad \varphi(\rstar)=\frac{r_{\rm ref}}{r_{\rm ref}+\rstar}.
  \label{eq:bandlaw_exact}
\end{equation}
For the centimetre-scale Ku and Ka kernels $\rstar\ll r_{\rm ref}$, so
$\varphi\simeq1$. The slope then equals $-b$ when $b\ll m$ and is mildly steeper
otherwise. A single-band profile does not fix the scale dependence of
$\Dtwo(r)$ by itself, so in practice $b$ is constrained by the dual-frequency data.
How well the band pair resolves $b$ is quantified in Section~\ref{sec:disc-ident}.

Two limits of \eqref{eq:marginallog} bound its use. First, it is a local,
centimetre-scale form. As $r\to\infty$ it grows as
$\tfrac12 ar^2+\tfrac12 b\,r_{\rm ref}r$ instead of saturating, so it is only the
small-lag asymptote of a structure function that saturates at large lag. This
growth does not affect the cross-section. A saturating continuation would act only
where $\Dtwo$ approaches $2\sigma_\eta^2$, and there the integrand weight
$e^{-\qz^2\sigma_\eta^2}$ is negligible, since $\qz^2\sigma_\eta^2\gtrsim10^3$.
Beyond about $0.14$\,m, moreover, the fitted $s^2(r)$ no longer increases with
wind, so the form is used only at centimetre lags.

Second, the form allows the wind dependence to be carried either by the level
$a(\Uten)$ or by the log-slope $b(\Uten)$. At lags $r\ll r_{\rm ref}$ only the log-slope
$b\simeq-\partial s^2/\partial\ln r$ is independent of the reference, whereas the split between $a$ and $b$
depends on the arbitrary reference lag $r_{\rm ref}$. The wind dependence of $b$ is
therefore the physical quantity, and its origin is the subject of the next section.

\section{The Short-Wave Saturation Origin of $b(\Uten)$}
\label{sec:bU-origin}

The wind dependence carried by $b$ is traced here to the saturation level of the
short-gravity spectrum. To test this reading, the structure function is generated
instead of fitted: a physical short-wave spectrum is fed to the same Hankel
integral \eqref{eq:hankel}. We use the directional spectrum of
\citet{kudryavtsev2014}, hereafter KMC, whose short-wave range includes
wave-breaking and parasitic-capillary terms. The generated model reproduces the
measured near-nadir profile to $0.113\dB$, against $0.105\dB$ for the joint
retrieval. Profile agreement alone, however, does not test the log-slope, so the
comparison below is made on the amplitude of $b$.

\begin{figure*}[t]
  \centering
  \includegraphics[width=\textwidth]{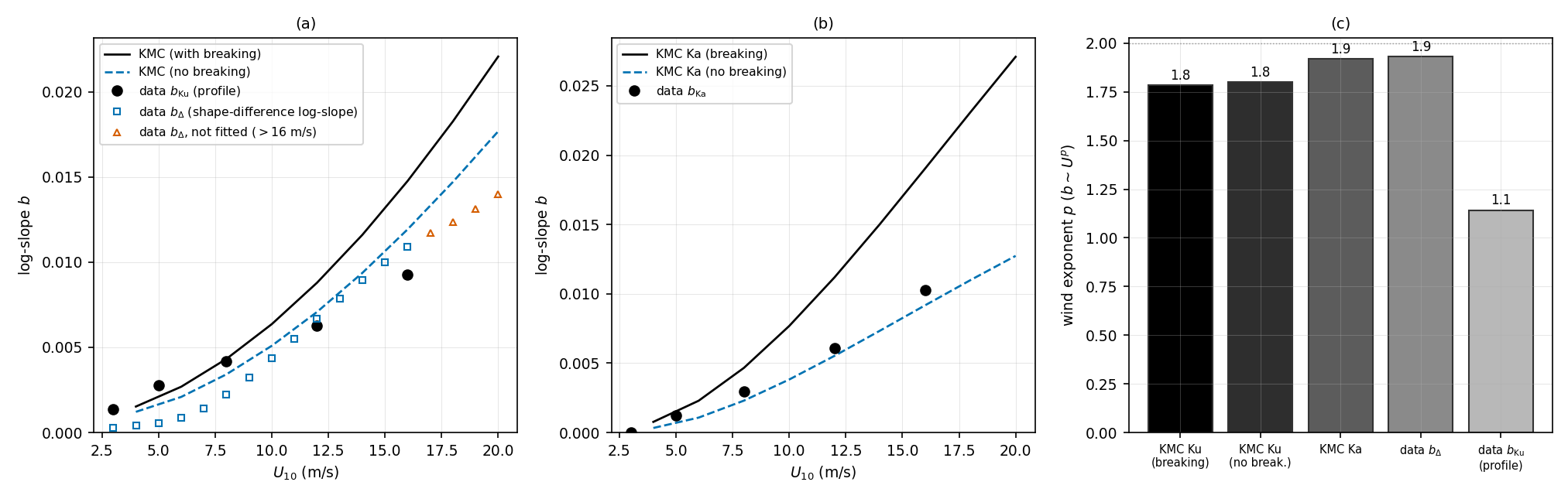}
  \caption{Where KMC over-predicts $b$. (a)~Ku and (b)~Ka: the KMC local
  log-slope with the breaking and parasitic terms included (solid) and excluded
  (dashed); filled markers: single-band fits $b_{\rm Ku/Ka}$; open markers in (a):
  measured shape-difference log-slope $b_\Delta$ (above $16\,\mathrm{m\,s^{-1}}$
  excluded from the joint fit).
  With breaking, KMC exceeds $b_\Delta$ by $35\%$ (Ku) and $74\%$ (Ka) at
  $16\,\mathrm{m\,s^{-1}}$. Without breaking, the Ka curve falls below the profile
  fit at the highest wind. (c)~Wind exponent $p$ of $b\propto\Uten^{\,p}$, fitted
  over $8$--$20\,\mathrm{m\,s^{-1}}$. KMC gives $p=1.8$ (Ku) and $1.9$ (Ka), weakly
  dependent on breaking, matching $b_\Delta$; the profile fit $b_{\rm Ku}$ gives a
  flatter $p\simeq1.1$, as it also absorbs the incidence shape.}
  \label{fig:overgrowth}
\end{figure*}

This comparison needs a physical reading of $b$, which the short-wave spectrum
supplies. The quantity $s^2(r)\equiv2\Dtwo(r)/r^2$ approximates the mean-square
slope of the waves longer than $r$, $\mss(k<1/r)$. Consider an elevation spectrum
$E(k)=b\,k^{-3}$, that is, a constant saturation spectrum $B(k)=k^3E(k)=b$. Its
slope spectrum $k^2E(k)=b/k$ gives $\mss(k<1/r)=a+b\ln(1/r)$, so every factor of ten
in wavenumber adds the same mean-square slope, $b\ln10$. This is the $k^{-3}$
saturation range of \citet{phillips1958}, distinct from the $k^{-5/2}$ equilibrium
range of \citet{phillips1985}, and $b$ is its saturation level. A synthetic
$k^{-3}$ spectrum confirms this reading: its structure function is
indistinguishable from \eqref{eq:marginallog}, and the fit returns the input level
to within $2\%$. The logarithmic form is thus the structure function of a
saturation range. Retrieved from the radar data, $b$ grows approximately as the
square of the wind speed (Fig.~\ref{fig:overgrowth}c). This is faster than the
short-wave level of the \citet{elfouhaily1997} spectrum, which grows only
logarithmically with wind. Generating $b$ from the KMC spectrum shows where the
faster growth comes from.

Before that comparison, the observed $b$ must be shown to be a property of the
surface.\footnote{\emph{Data used in the diagnostic figures.}
Figs.~\ref{fig:overgrowth}, \ref{fig:boundwave} and \ref{fig:go4} use three measured
quantities. The first is the 2021 azimuth-averaged $\snot(\theta,\Uten)$ of the GPM
Ku-band precipitation radar (KuPR, V07A), binned at $0.25^\circ$ and
$0.5\,\mathrm{m\,s^{-1}}$ of ERA5 reanalysis $10$-m wind for the reference sea state,
significant wave height $\Hs\in[1.5,3.5]$\,m ($902$ bins over $0$--$18^\circ$). The
second is the matched GPM shape difference
$\Delta\mathrm{mss_{shape}}=\mathrm{mss_{shape}^{Ka}}-\mathrm{mss_{shape}^{Ku}}$. In each
band, $\mathrm{mss_{shape}}=-10/(\ln10\,\beta)$, where $\beta$ is the slope of a
regression of $10\log_{10}(\cos^4\theta\,\snot)$ on $\tan^2\theta$ over
$2$--$9^\circ$, from 19 months (2021-12 to 2023-07). The
third is the set of CFOSAT/SWIM azimuth harmonics $a_1,a_2$. Residuals are weighted
by $w_i=\sqrt{n_i}$ ($n_i\le10^7$, unit mean), and the weighted root-mean-square
difference is $\mathrm{wRMSD}=\big(\sum w_ie_i^2/\sum w_i\big)^{1/2}$.} Fitted independently in each band, it agrees between Ku and Ka above about
$10\,\mathrm{m\,s^{-1}}$, to within $3\%$ at $12$ and $10\%$ at
$16\,\mathrm{m\,s^{-1}}$ (Fig.~\ref{fig:overgrowth}). This agreement supports a
common surface interpretation. At low wind the single-band Ka estimate
collapses toward its bound, because the Ka profile then carries almost no
information on the log-slope.

KMC reproduces the wind scaling of this $b$. Over $8$--$20\,\mathrm{m\,s^{-1}}$, a
power law fitted to the generated $b$ gives an exponent $p\simeq1.8$ at the Ku and
$\simeq1.9$ at the Ka kernel lag, with or without the breaking and
parasitic-capillary terms ($1.78$--$1.80$ and $1.86$--$1.92$). The wind exponent
therefore comes from the base saturation of KMC,
$B(k)\propto[c_\beta(u_*/c)^2]^{1/n}$, where $u_*$ is the friction velocity, $c$
the phase speed, $c_\beta$ the wind-growth constant and $n$ the dissipation
exponent, which tends to $1$ near the gravity--capillary wavenumber $k_\gamma$. The
same ${\approx}\Uten^{2}$ scaling is found independently in the dual-copolarized
Ka-band model of \citet{yurovsky2017}, whose non-polarized component scales as
about $\Uten^{2}$.

The breaking and parasitic terms, in contrast, mainly raise the level of $b$, and
they account for most of its high-wind excess. At $16\,\mathrm{m\,s^{-1}}$ the
generated $b$ lies $35\%$
above the shape-difference log-slope $b_\Delta$ at the Ku kernel lag and $74\%$
above it at Ka. The excess is larger at Ka because the Ka kernel wavenumber
$1/\rstar$ lies closer to $k_\gamma$, where the small-scale wind growth is
steepest. In absolute terms the excess grows with wind. In relative terms it is
largest at low wind, where $b_\Delta$ is small and least certain.

This excess points to the short-wave spectrum rather than to the tested slope
probability density functions (PDFs). The short-wave saturation of KMC is constrained
mainly by slope statistics, namely the \citet{cox1954} mean-square slope and
capillary-wave spectra, as the roughness synthesis of \citet{hwang2022} also shows.
The Ku--Ka shape difference weights a different quantity, the slope variance
between the two kernel wavenumbers, that is, the partial moment $\int B\,d\ln k$. It
is not the mean-square curvature $\mathrm{msc}=\int k^4E\,dk$, which is carried by
still shorter capillaries. KMC thus lies only $7.5\%$ above the Cox--Munk total slope variance
and is still $35\%$ (Ku) and $74\%$ (Ka) too high over the
centimetre band that the radar weights.

No tested positive increment density, whether Gaussian, maximum-entropy or the
Gaussian scale mixture of \citet{tatarskii1999b,tatarskii2003}, removes the
short-wave excess.
\citet{tatarskii1999b} already pointed out this tension between the spectrum and
the slope PDF. Fitting $\Dtwo(r)$ jointly to the profile and shape difference
reduces this tension.

A per-wind diagnostic fit keeps a close incidence-profile fit while matching the shape
difference.
For the logarithmic form, the Ku--Ka shape difference is controlled mainly by
$b\,\Delta L$, where $\Delta L$ is the $\ln r$ interval between the two kernel lags,
fixed by the band geometry. When $a$, $b$ and a level are fitted in each wind bin to the
$0$--$11^\circ$ KuPR profile and to the shape difference, weighted $200$ times the
dB profile residuals, the strongly weighted shape
difference is met by construction. The profile then keeps a wRMSD of
$0.060$--$0.068\dB$ per wind bin over $0$--$11^\circ$ and $6$--$14\,\mathrm{m\,s^{-1}}$.
No additional change of slope, or knee, is needed in $\Dtwo$: adding one lowers the
profile wRMSD by less than $0.001\dB$ and drives $b$ to its lower bound.

The generated spectrum thus supplies the logarithmic form of $\Dtwo$ and its wind
exponent, about $\Uten^{1.8\text{--}1.9}$, while the radar measures the high-wind
amplitude. The data-fitted logarithmic form is adopted in the physical core.

\section{Higher Cumulants as Optional Extensions}
\label{sec:d4}

The core model needs only $\Dtwo$, so an independently fitted fourth cumulant is an
optional extension, which KuKa-Mod does not use (Table~\ref{tab:hier}). Physical
fourth cumulants need not vanish. Once the band pair fixes $\Dtwo$, a residual
profile shape could be read either as further scale dependence of $\Dtwo$ or as a
non-Gaussian increment, and the two bands do not separate them. Variability of the
local variance also produces kurtosis. For increments that are zero-mean Gaussian
conditional on a state $U$, $D_4=3\,\mathrm{Var}_U[\Dtwo(r;U)]$. A wind that varies
within a footprint, or among collocations assigned the same reference wind, therefore
contributes a $D_4$ whenever the conditional variance varies, and
\citet{coxmunk1954,coxmunk1956} already traced the peakedness of their glitter slopes
possibly to wind-speed variability, a standard deviation of about one third of the
mean wind being enough to produce it. The same holds for a footprint $\Hs$ at fixed
wind, when the increment stays zero-mean Gaussian given $\Hs$: the tilt of
Section~\ref{sec:disc-hs} makes $\Dtwo(r;\Hs)$ depend on $\Hs$, and its spread
contributes $D_4=3\,\mathrm{Var}_{H_s}[\Dtwo(r;\Hs)]$. The electromagnetic sea-state
bias is a backscatter--elevation correlation carried by the long waves
\citep{osina2026}; that correlation alone does not imply a nonzero $D_4$. A Gaussian fitted core does not establish Gaussian surface
statistics.

If a fourth cumulant is admitted, it should keep the characteristic function
normalizable. Following \citet{chapron2000}, slope peakedness is read as the
signature of a slope field whose local variance fluctuates, along the longer waves
or with the wind.
Let $\tilde s^2$ be that local radar-band slope variance divided by its mean, and
let it follow a Gamma law with relative variance $R=\mathrm{Var}(\tilde s^2)$. The
standardized two-point characteristic function then reads
\begin{equation}
  \hat\chi(t)=\Big(1+\tfrac12 R\,t^2\Big)^{-1/R},\qquad
  \text{excess kurtosis}=3R,
  \label{eq:kdist}
\end{equation}
where $t$ is the dimensionless argument, evaluated as $\qz\sqrt{\Dtwo}$ in the
scattering integral. This K-type slope model is positive-definite and normalizable
for every $R>0$, unlike a truncated Gram--Charlier polynomial, and $R\to0$ recovers
the Gaussian model. With the non-saturating $\Dtwo\sim ar^2/2$, however, the nadir
scattering integral converges only for $R<1$, its integrand scaling as $r^{1-2/R}$. Its expansion fixes the whole even hierarchy, $D_4=3R\Dtwo^2$
and $D_6=30R^2\Dtwo^3$. The two parameters act differently, since $b$ shapes
$\Dtwo$ whereas $R$ acts on the increment density at all lags. The present data
nevertheless do not determine them independently (Section~\ref{sec:disc-ident}).

The Gamma law is a normalizable proxy for a physical mixing law. Choppy-wave
straining \citep{creamer1989,nouguier2009} gives the local slope variance
$s_L^2=\mss_0(1+e_L)^{-2}$, where $\mss_0$ is the unstrained slope variance and
$e_L$ the long-wave strain, of variance $\sigma^2_{\rm long}$. Both laws share the
same leading relative variance, $R\simeq4\sigma^2_{\rm long}$.

\begin{figure*}[t]
  \centering
  \includegraphics[width=\textwidth]{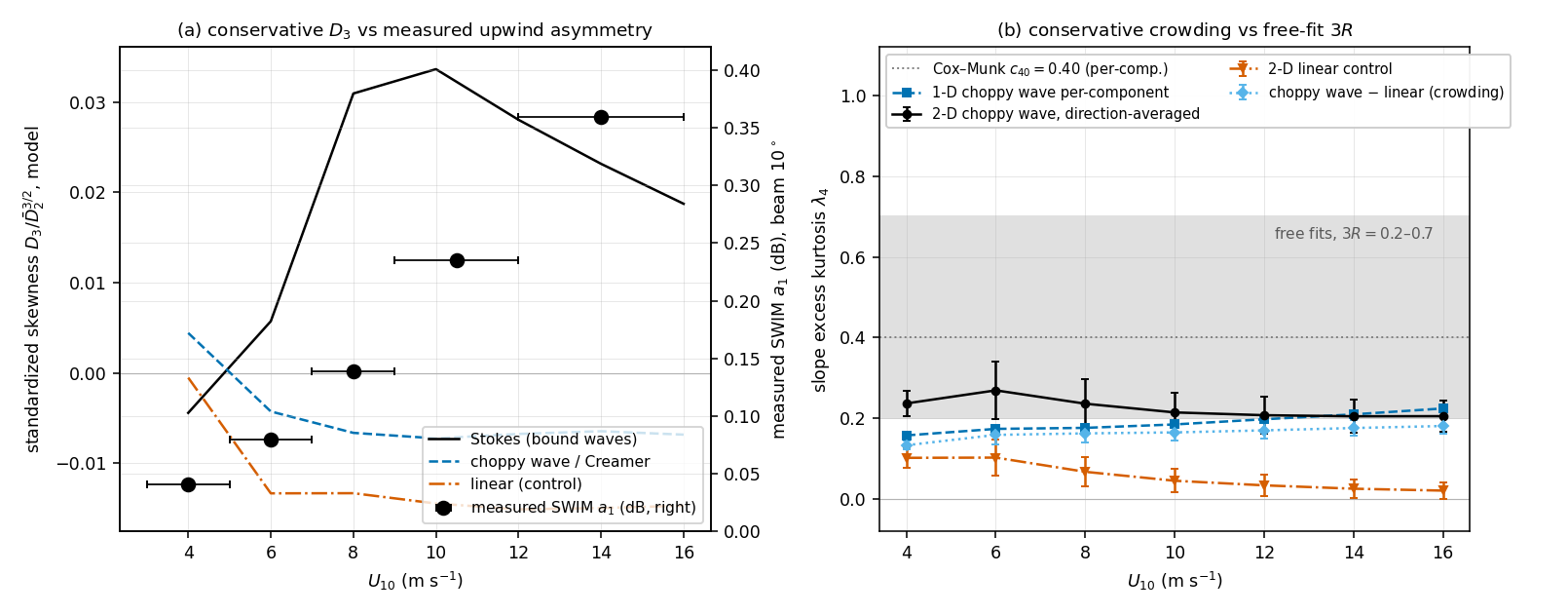}
  \caption{Filtered-slope cumulants ($k_d=k_r/3$) as proxies for the increment
  cumulants at the radar lags, from the \citet{kudryavtsev2014} spectrum:
  one-dimensional ensembles in (a), $1024^2$ realizations in a $20$\,m box with eight
  seeds in (b). (a)~Third cumulant: the
  Eulerian and choppy-wave second-order surfaces give a standardized slope skewness
  of at most $0.034$ (left axis, with the linear control); the SWIM $a_1$ at
  $10^\circ$ (right axis) reaches $0.36\dB$.
  (b)~Fourth cumulant. Circles are the direction-averaged excess kurtosis of the
  slope components of the choppy-wave surface, triangles its linear
  anisotropic-Gaussian control, and squares the per-component one-dimensional
  estimate. The difference between circles and triangles, the conservative
  crowding contribution (diamonds), is $0.13$--$0.18$ (mean $0.16$) and nearly
  independent of wind. The shaded band is the envelope $3R=0.2$--$0.7$ of free
  fits, not a data requirement, and the dotted line the Cox--Munk per-component
  $c_{40}=0.40$. Bars: eight-seed spread (b), wind-bin widths (a).}
  \label{fig:boundwave}
\end{figure*}

Filtered-slope proxies were computed from KMC surfaces with the cutoff
$k_d=k_r/3\simeq94.8\,\mathrm{rad\,m^{-1}}$ (Fig.~\ref{fig:boundwave}), so that
$k_d\rstar\simeq1.3$--$2.4$ at Ku. Choppy-wave crowding raises the direction-averaged
slope excess kurtosis by $0.13$--$0.18$ (mean $0.16$) above its anisotropic-Gaussian
control over $4$--$16\,\mathrm{m\,s^{-1}}$, nearly independently of wind; finer
($1536^2/20$\,m) and larger ($2048^2/40$\,m) grids change it by at most $0.017$.
These checks do not establish finite-lag increment convergence. Resonant four-wave interactions add at most
about $0.025$ \citep{janssen2009}.

The second source, long-wave modulation of the radar-band mean-square slope, would
supply the part that rises with wind,
\begin{equation}
  R=|M|^2\,\sigma^2_{\rm long},\qquad
  \sigma^2_{\rm long}=\!\!\int_{k<3\,{\rm rad\,m^{-1}}}\!\! k^2 E(k)\,dk.
  \label{eq:Rmod}
\end{equation}
Here $M$ is the hydrodynamic modulation transfer function (MTF), and the long-wave
slope variance $\sigma^2_{\rm long}$ rises from $0.008$ to $0.021$ over
$4$--$16\,\mathrm{m\,s^{-1}}$. The MTF can be estimated from wave-action
conservation. When a long wave compresses each short-wave packet as
$k\to k(1+\delta)$, neglecting the change of surface area, the straining response
is $m_s=d\ln\mss/d\delta=5/2$ \citep{keller1975,alpers1978}. With relaxation at the
rate $\mu$ of \citet{plant1982}, evaluated at $k=1/\rstar$, the mean-square-slope
MTF per unit long-wave slope becomes
\begin{equation}
  |M|=m_s\,\frac{\Omega}{\sqrt{\mu^2+\Omega^2}},
  \label{eq:Mpred}
\end{equation}
where $\Omega\simeq2.4$--$3.6\,\mathrm{rad\,s^{-1}}$ is the slope-weighted
frequency $\sqrt{gk}$ of the long waves. At Ku, $|M|$ falls from $2.5$ at $4$ to
$1.9$ at $16\,\mathrm{m\,s^{-1}}$, the same order as the $|M|\simeq1.5$ fitted to
the wind trend of the kurtosis. Because this estimate rests on simplified
conservation and transport assumptions \citep{peureux2021,curcic2025}, it sets the
order of $R$ but not its value.

\section{The Third Cumulant $D_3$}
\label{sec:d3}

Whereas the fourth cumulant controls the peakedness, the odd cumulant $D_3$ is the
source of the upwind/downwind asymmetry of $\snot$.
Because $D_3$ changes sign under lag reversal $\bm r\to-\bm r$, it enters the
exponent of \eqref{eq:cumulants} as an imaginary phase,
$-\tfrac{i}{6}\qz^3D_3(\bm r)$, and produces the $\cos\Phi$ azimuth harmonic $a_1$.
SWIM measures this harmonic at amplitudes up to $0.36\dB$ on the $10^\circ$ beam.

To model it, $D_3$ needs a closure that ties it to the even structure function. We
write $D_3=\lambda_3(\Uten)\,D_{20}^{\,\nu}\cos\psi$, where $\psi$ is the azimuth
of the lag relative to the wind and $D_{20}(r)$ is the azimuth-averaged part of
$\Dtwo$. This form vanishes at zero lag, as any cumulant of an increment must. The
strength of the asymmetry is best measured by the dimensionless standardized
skewness $D_3/D_{20}^{3/2}$.

The exponent $\nu$ is chosen from the data. The scale-invariant choice
$\nu=\tfrac32$ keeps the standardized skewness the same at all lags, as the
Cox--Munk slope skewness used by \citet{mouche2007} implies. It makes $a_1(\theta)$
turn over near $9^\circ$, whereas the measured SWIM $a_1$ rises monotonically up to
the outermost beam. The linear choice $\nu=1$ keeps $a_1$ rising and matches the
data at all beams.

The linear choice, however, cannot hold down to zero lag. With $\nu=1$ the
standardized skewness grows as $D_{20}^{-1/2}$ when $r\to0$ and eventually
exceeds the Pearson bound, which limits the skewness allowed for a given kurtosis.
The value $\nu=1$ is therefore an effective exponent over the radar lags only. A
regularized form keeps it there and restores a finite skewness at small lags:
\begin{equation}
  D_3(r,\psi)=\lambda_3^{\,0}\,
  \frac{D_{20}^{3/2}}{\sqrt{1+D_{20}/D_c}}\,\cos\psi.
  \label{eq:d3reg}
\end{equation}
Here $D_c$ is a crossover increment variance. Where $D_{20}\gg D_c$,
\eqref{eq:d3reg} reduces to the linear closure, and where $D_{20}\ll D_c$ it
returns to the scale-invariant $\nu=\tfrac32$. With $D_c=10^{-7}$\,m$^2$ the fitted
amplitudes stay below $1.24$, inside the Pearson bound $\sqrt2$ at Gaussian kurtosis,
down to zero lag. Over $\rstar/2\le r\le2\rstar$ at the fitted winds, $D_{20}/D_c$
exceeds $2$ at Ka and $10$ at Ku, so the radar sees a nearly linear regime.
The data accordingly constrain mainly the product $\lambda_3^{0}\sqrt{D_c}$, the slope of
the linear closure over the radar lags.

Conservative wave nonlinearity is a weak candidate for this skewness. Second-order
bound waves give only a small slope skewness, as \citet{longuet1982} showed, and
simulated KMC surfaces confirm it. Both the Eulerian and the Lagrangian
(choppy-wave) second-order surfaces give a standardized slope skewness of at most
$0.034$, a filtered-slope proxy with a large sampling uncertainty
(Fig.~\ref{fig:boundwave}a). The fitted closure reaches $0.10$--$0.16$ at the Ku
kernel lag over $10$--$14\,\mathrm{m\,s^{-1}}$. Comparing the two in backscatter
needs a common forward calculation. Skewness can arise from correlations between
variance and slope. One source is the modulation of the short waves by the long
waves: short waves are steeper on some phases of a long wave, coherently with the
wind, as \citet{chapron2002} found for steep wavelets and near-breaking waves. Such
correlations can also act at all scales, as in the crescent (horseshoe) patterns that
developing waves form \citep{su1982}, which contribute to $a_1$ only through a
wind-aligned fore--aft asymmetry, since exact statistical isotropy gives $D_3=0$.
$D_3$ therefore does not select a unique scale partition.

The law of total cumulance links $D_3$ and $D_4$. Write the elevation increment over
a radar lag as $X=L+\epsilon$, where $L=\mathrm{E}[X\,|\,H]$ is centered and
$\epsilon\,|\,H\sim\mathcal N(0,V(H))$ for a conditioning state $H$. Then
\begin{equation}
  \begin{aligned}
  \kappa_3(X)&=\kappa_3(L)+3\,\mathrm{Cov}(L,V),\\
  \kappa_4(X)&=\kappa_4(L)+6\,\mathrm{Cov}(L^2,V)+3\,\mathrm{Var}(V).
  \end{aligned}
  \label{eq:totalcum}
\end{equation}
A long-wave state is one choice of $H$. For the directional slope
$s=\hat{\bm r}\cdot\nabla\eta$, $3\,\mathrm{Cov}(L,V)\sim3r^3\,
\mathrm{Cov}(\mathrm{E}[s|H],\mathrm{Var}(s|H))$ as $r\to0$, the variance--slope part of
the $D_3$ of this section, $\kappa_3(L)$ being the other.
The same variability contributes to $D_4$ through $\mathrm{Var}(V)$ and
$\mathrm{Cov}(L^2,V)$ (Section~\ref{sec:d4}). Predicting both amplitudes requires
these joint statistics, so the kurtosis is estimated from $|M|$ whereas $\lambda_3$
is still taken from data.

\section{The Complete Model and Its Ku/Ka Instantiation}
\label{sec:complete}

The three cumulant terms can now be assembled into one GMF. Its azimuth-averaged
part is the Kirchhoff--Hankel integral of a single two-point characteristic
function, the directional terms enter as azimuth harmonics, and an empirical Bragg
term extends the model to the outer incidence flank:
\begin{equation}
  \begin{aligned}
  \snot(\theta,\Uten,\Phi)
  ={}& \bar\snot(\theta,\Uten)\,
    \big[1 + A_1\cos\Phi + A_2\cos2\Phi\big]\\
    &{}+ \snot_{\rm Bragg},
  \end{aligned}
  \label{eq:full}
\end{equation}
Here the azimuth-averaged specular part is
\begin{equation}
  \begin{aligned}
  \bar\snot(\theta,\Uten)
  &= \frac{2k_r^2 R_{\rm eff}^2}{\cos^2\theta}
    \int_0^\infty\! r\,J_0(\qh r)\\
  &\quad\times\,
    \hat\chi\!\big(\qz\sqrt{\Dtwo(r,\Uten)};\,R(\Uten)\big)\,dr,
  \end{aligned}
  \label{eq:full_mean}
\end{equation}
and the three structure-function terms are
\begin{align}
  \text{($\Dtwo$)}\;\;
  &\Dtwo(r,\Uten)=\tfrac12 r^2\big[a(\Uten)+b(\Uten)\ln(1+r_{\rm ref}/r)\big],
  \label{eq:full_d2}\\[1mm]
  \text{($D_4$)}\;\;
  &\hat\chi(t;R)=\big(1+\tfrac12 R\,t^2\big)^{-1/R},\notag\\
  &R(\Uten)=|M|^2\sigma^2_{\rm long}(\Uten),
  \label{eq:full_chi}\\[1mm]
  \text{($D_3$)}\;\;
  &A_1\ \text{from the regularized}\ D_3(r,\psi)\ \text{of \eqref{eq:d3reg}}.
  \label{eq:full_d3}
\end{align}
The second harmonic $A_2$ comes from the $\cos2\psi$ anisotropy of $\Dtwo$,
$D_{22}=\tfrac12 r^2[\delta_a+\delta_b\ln(1+r_{\rm ref}/r)]$, whose coefficients
$\delta_a,\delta_b$ are listed in Table~\ref{tab:params}. The harmonic bracket
$[1+A_1\cos\Phi+A_2\cos2\Phi]$ in \eqref{eq:full} is the first-order representation
of the anisotropic cumulant integral. It collects the leading Fourier harmonics
generated by the anisotropic $\Dtwo$ and the odd $D_3$ in the exponent of
\eqref{eq:cumulants}. For the quadratic ($\delta_a$) part of $D_{22}$ it coincides
with the first-order expansion of the elliptical GO law
(Section~\ref{sec:golimit}). The harmonics are quoted in decibels as
$a_n\!\simeq\!(10/\ln10)A_n$. First order is adequate, because the observed SWIM
first harmonic, several tenths of a decibel, is of order ten percent in linear
power.

The Bragg term $\snot_{\rm Bragg}$, in contrast, lies outside the specular theory.
It is an empirical first-order small-perturbation (SPM-1) Bragg term, switched on
from about $10^\circ$ and added incoherently. It supplements the specular integral on
the outer flank, extending KuKa-Mod to $18^\circ$. The specular core
itself is used over $0$--$11^\circ$.

The assembled model keeps the limits established above. Equation
\eqref{eq:full_chi} reduces to the Gaussian model as $R\to0$, and the isotropic specular core of \eqref{eq:full}
reduces to Gaussian, isotropic geometric optics when, in addition, $b\to0$. Wind enters the core through
$a(\Uten)$ and $b(\Uten)$, of which only $b$ is independent of $r_{\rm ref}$. The
optional extensions add $R(\Uten)$, $\lambda_3(\Uten)$ and the directional
anisotropy of $\Dtwo$. The radar band enters through $k_r$ and $R_{\rm eff}^2$, and
sea state through the long-wave anomaly $\Delta\sigma^2_{\rm long}(\Uten,\Hs)$ of Section~\ref{sec:disc-hs}.

These terms combine into the configurations of Table~\ref{tab:hier}, from the
quadratic GO limit $\mathrm{M_0}$ through the core $\mathrm{M_{joint}}$ to the optional
and spectrum-driven ones. Table~\ref{tab:params} lists the parameters with their
physical origin, and Table~\ref{tab:ledger} separates what the theory derives from what
the model still fits or calibrates, which defines the open problems of
Section~\ref{sec:discussion}.

The operational Ku/Ka realization, \emph{KuKa-Mod}, combines the core
configuration $\mathrm{M_{joint}}$ with the directional extension. Its core has
no independently fitted kurtosis ($R=0$), and it
references the Ka level to the dielectric ratio and the measured attenuation, up to
a residual calibration term. Its fitting, validation and release are the subject of
Part~II.

Equations \eqref{eq:full}--\eqref{eq:full_d3} describe the return for one local
wind, whereas a radar cell averages over many. A near-nadir cell spans hundreds of
metres to kilometres, across which the $10$-m wind is not homogeneous. Because
$\snot$ is nonlinear in $\Dtwo$, the cell mean is the Kirchhoff kernel averaged over
the wind within the cell, of mean $\bar\Uten$ and variance $\sigma_U^2$. This average
follows the same scale-mixture algebra as \eqref{eq:kdist}, with the cell wind as
the mixing variable. To order $\sigma_U^2$ it modifies the structure function into
\begin{equation}
  \Dtwo^{\rm eff}(r)=\Dtwo(r;\bar\Uten)
   +\tfrac12\,\partial_U^2\Dtwo\,\sigma_U^2
   -\tfrac14\,q_z^2\,(\partial_U\Dtwo)^2\,\sigma_U^2 ,
  \label{eq:d2eff}
\end{equation}
where $\partial_U$ denotes the derivative with respect to $\Uten$, valid while
$\tfrac14 q_z^2(\partial_U\Dtwo)^2\sigma_U^2\ll\Dtwo$. The first correction is a
Jensen term that raises the saturation level, and the second is the image in
$\Dtwo^{\rm eff}$ of the wind-mixture kurtosis $D_4\simeq3(\partial_U\Dtwo)^2\sigma_U^2$. The error of a
reference wind produces the same mixing among collocations. Because $\Dtwo^{\rm eff}$ depends on $q_z$, it is an
effective quantity. We prescribe the gustiness
$\sigma_U=[0.5^2+(0.10\,\Uten)^2]^{1/2}(\ell/17.5\,\mathrm{km})^{1/3}$ (winds in
$\mathrm{m\,s^{-1}}$), where $\ell$ is the footprint size, and evaluate the
correction by Gaussian wind averaging, with no amplitude fitted. It matters only for
the coarse footprints of real-aperture near-nadir radars.

\begin{table*}[t]
  \centering
  \caption{Configurations of the structure-function Kirchhoff framework. The
  dual-frequency-constrained configuration $\mathrm{M_{joint}}$ is the core model.
  The extensions $\mathrm{M_R}$ and $\mathrm{M_{dir}}$ have empirical amplitudes, and
  the spectrum-driven configuration $\mathrm{M_{spec}}$ serves as a physical prior.
  The ``Fitted'' column lists only the quantities that carry the wind dependence.
  All configurations share the level prefactor $R_F^2 C_f$
  (Table~\ref{tab:params}). The Bragg flank is not part of these specular
  configurations, but an operational addition of KuKa-Mod.}
  \label{tab:hier}
  \footnotesize
  \setlength{\tabcolsep}{4pt}
  \renewcommand{\arraystretch}{1.25}
  \begin{tabular}{@{}p{0.07\textwidth}p{0.34\textwidth}p{0.19\textwidth}p{0.20\textwidth}p{0.08\textwidth}@{}}
    \toprule
    Config. & Definition & Fitted (wind-dependent) & Constrained by & Role \\
    \midrule
    $\mathrm{M_0}$ & Gaussian, isotropic GO: quadratic $\Dtwo$ \eqref{eq:go} &
      $\mss(\Uten)$ & --- (analytic limit) & baseline \\
    $\mathrm{M_{joint}}$ & logarithmic $\Dtwo$, Gaussian increment
      \eqref{eq:full_d2}; \textbf{core} &
      $a(\Uten),b(\Uten)$ & profile $+$ matched dual-freq.\ shape diff. & core \\
    $\mathrm{M_R}$ & $\mathrm{M_{joint}}$ $+$ variance-mixture kurtosis $R$
      \eqref{eq:full_chi} & $+\,R(\Uten)$ & profile $+$ band & extension \\
    $\mathrm{M_{dir}}$ & $\mathrm{M_{joint}}$ $+$ $\cos2\psi$ anisotropy $+$ odd
      $D_3$ \eqref{eq:full_d3} & $+\,\delta_a,\delta_b,\lambda_3^{0}$ &
      azimuth harmonics & extension \\
    $\mathrm{M_{spec}}$ & spectrum/MTF-driven predictive form &
      none (spectrum-prescribed) & profile (no band) & prior \\
    \bottomrule
  \end{tabular}
\end{table*}

\begin{table*}[t]
  \centering
  \caption{Parameters of the model \eqref{eq:full}, with their status and origin. The
  core $\mathrm{M_{joint}}$ fits only $a(\Uten)$ and $b(\Uten)$; $R$, $\lambda_3$ and
  $\delta_{a,b}$ belong to the optional extensions of Table~\ref{tab:hier}, and the three
  Bragg parameters to the operational flank.}
  \label{tab:params}
  \footnotesize
  \setlength{\tabcolsep}{5pt}
  \renewcommand{\arraystretch}{1.25}
  \begin{tabular}{@{}p{0.10\textwidth}p{0.07\textwidth}p{0.36\textwidth}p{0.14\textwidth}p{0.21\textwidth}@{}}
    \toprule
    symbol & term & meaning & status & value / origin \\
    \midrule
    $R_F^2(f,T,S)$ & prefactor & physical nadir Fresnel power reflectivity &
      computed & $0.616$ (Ku, $18.3^\circ$C, $S=35$); dielectric model \\
    $C_f$ & prefactor & per-band calibration, path and nugget factor (Ka absorbs residual
      atmospheric attenuation) & calibrated & band-specific \\
    $a(\Uten)$ & $\Dtwo$ & additive coefficient of \eqref{eq:full_d2} at
      $r_{\rm ref}$; not a physical plateau &
      fitted in $\mathrm{M_{joint}}$ & $0.002$--$0.024$ at the six wind nodes
      (joint retrieval) \\
    $b(\Uten)$ & $\Dtwo$ & small-lag log-slope of $s^2(r)$; sets the
      dual-frequency shape difference & fitted; spectrum supplies prior &
      $0$--$0.014$ (joint retrieval; single-band profile fits inflate the
      low-wind values) \\
    $r_{\rm ref}$ & $\Dtwo$ & reference scale (sets the $a$/$b$ split) &
      fixed & $1$\,m (convention) \\
    $R(\Uten)$ & $D_4$ & variance-mixture kurtosis ($3R$);
      long-wave estimate $R\simeq|M|^2\sigma^2_{\rm long}$, $|M|=\tfrac52\Omega/\sqrt{\mu^2{+}\Omega^2}$ &
      optional; $|M|$ estimated & $\sigma^2_{\rm long}$ from the long-wave spectrum \\
    $\lambda_3^{0}$ & $D_3$ & odd-harmonic amplitude (regularized closure \eqref{eq:d3reg}) &
      optional; fitted & amplitude from data; form from SWIM \\
    $\delta_a,\delta_b$ & $\Dtwo$ anis. & up/crosswind $\Dtwo$ anisotropy ($a_2$) &
      optional; fitted & from data, offset-removed \\
    $(\beta,\xi,\alpha)_{\rm Bragg}$ & flank & gated SPM-1 Bragg add-on
      ($11$--$18^\circ$) & empirical & operational extension \\
    \bottomrule
  \end{tabular}
\end{table*}

\begin{table}[t]
  \centering
  \caption{Where the theory predicts directly, and where the model still relies on
  data. The boundary marks the open theoretical problems
  (Section~\ref{sec:discussion}).}
  \label{tab:ledger}
  \footnotesize
  \setlength{\tabcolsep}{3pt}
  \begin{tabular}{p{0.30\columnwidth}p{0.62\columnwidth}}
    \toprule
    \textbf{Derived from theory} & \textbf{Still taken from data} \\
    \midrule
    Kirchhoff--Hankel master eq.; exact GO limit; kernel lag $\rstar$ &
      high-wind \emph{amplitude} of $b(\Uten)$: the source spectrum over-predicts the centimetre-scale tail \\
    logarithmic \emph{form} of $\Dtwo$; band-transfer law; conservative
      (choppy-wave) filtered-slope kurtosis proxy $0.13$--$0.18$ &
      long-wave contribution to $R$: short-wave relaxation rate $\mu$ not yet derived \\
    scale-mixture closure $\hat\chi$ (positive-definite Gamma proxy) &
      long-wave contribution to $\lambda_3$: MTF phase $\mathrm{Re}\,M$ not yet derived \\
    Fresnel ratio $R_F^2(f_2)/R_F^2(f_1)$ from the dielectric; order of the
      straining response, $m_s\simeq\tfrac52$ (estimate) &
      calibration/path/nugget factor $C_f$ (incl.\ Ka attenuation); sea-state coefficient
      $\partial\Delta\sigma^2_{\rm long}/\partial\Hs$ (uncalibrated, Part~II); gated Bragg flank ($11$--$18^\circ$ EM kernel) \\
    \bottomrule
  \end{tabular}
\end{table}

\section{Discussion}
\label{sec:discussion}

\subsection{Relation to the geometric-optics chain}
\label{sec:disc-go}
The logarithmic model and the classical geometric-optics chain share one forward
equation but represent scale dependence differently (Table~\ref{tab:concept}). The closest relative is the fourth-order model of \citet{nouguier2009phd}, published as
GO4 by \citet{boisot2015}. GO4 expands the same Kirchhoff integral through fourth order
in lag. For a smooth isotropic surface,
$\Dtwo(r)=\tfrac12\mss\,r^2-\tfrac1{32}\mathrm{msc}\,r^4+o(r^4)$. This $r^4$ term
belongs to the second cumulant, even for Gaussian increments, but it adds to the
statistical $D_4$ in the peakedness
$\alpha=\tfrac23\lambda_4+\mathrm{msc}/(q_z^2\mss^2)$ of \citet{nouguier2016}. The
curvature part depends on the radar frequency and vanishes as $\qz\to\infty$, so
that $\alpha\to\tfrac23\lambda_4$, the GO contribution from slope kurtosis. Retaining $\Dtwo$
over the lags that the radar weights therefore avoids a prescribed two-scale split of
the specular core.

For the logarithmic $\Dtwo$ an expansion in powers of
$r$ does not even exist, because $\Dtwo\propto r^2\ln r$ is not analytic at $r=0$,
and the logarithmic form does not define the optical limit. The GO4 curvature term
can thus be read as a finite-order approximation of the same scale dependence.

\begin{figure*}[t]
  \centering
  \includegraphics[width=\textwidth]{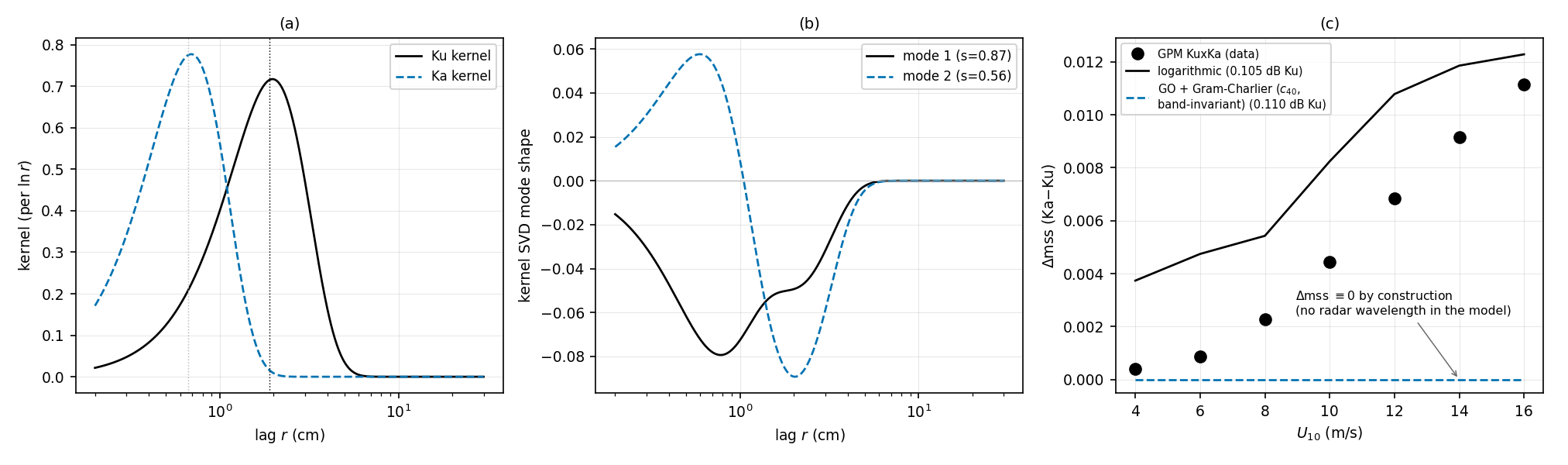}
  \caption{Single-band equivalence and complementary band kernels.
  (a)~Near-nadir contribution kernels per unit $\ln r$ for Ku and Ka, centred at
  $\rstar\simeq1.9$ and $0.7$\,cm (Ku-only fit parameters,
  $8\,\mathrm{m\,s^{-1}}$). (b)~Orthogonal combinations of the normalized
  kernels from a singular value decomposition (SVD; singular-value ratio $1.6$), which describe their overlap. This
  ratio is not the profile-Jacobian ratio of Section~\ref{sec:disc-ident}. (c)~Band transfer from a Ku-only fit. The logarithmic form predicts a shape
  difference with the correct sign but over-predicts it from $+10\%$ at $16$ to
  ninefold at $4\,\mathrm{m\,s^{-1}}$. The Gram--Charlier control ($16$ parameters)
  fits the Ku profile equally well but has $\Delta\mathrm{mss}\equiv0$ by construction.}
  \label{fig:go4}
\end{figure*}

Within one band this approximation works. Fitted to the same Ku profile over
$0$--$18^\circ$, geometric optics with six mss and six band-invariant Gram--Charlier
$c_{40}$ wind nodes and the logarithmic model fit equally well, with wRMSD of
$0.110$ and $0.105\dB$ (Fig.~\ref{fig:go4}), but this control has a zero Ku--Ka
difference by construction. The full GO4 predicts one once $\mathrm{msc}(\Uten)$ is
known; a Ku profile alone cannot split the total peakedness into kurtosis and
curvature, so predicting Ka requires a spectrum that supplies msc or a second
frequency.

The present method takes the second route. One $\Dtwo$ is evaluated at both
$\rstar_{\rm Ku}$ and $\rstar_{\rm Ka}$. Even fitted to Ku alone, it predicts a shape
difference that grows with wind with the correct sign (Fig.~\ref{fig:go4}c),
although too large at low wind (Section~\ref{sec:disc-ident}). Fitted to both bands,
it reproduces the matched difference. The new content is therefore the
dual-frequency inverse use: one centimetre-scale $\Dtwo(r)$ retrieved from matched
Ku$\times$Ka observations without an imported spectrum. The forward equation is
shared with GO4, and the one-band inversion for total mss and msc was already
performed by \citet{nouguier2009phd}.

Part~II compares the full GO4 with the logarithmic form on equal footing, with
$16$ parameters in each family. The two fit the Ku profile comparably, both need the
second frequency, and the logarithmic form transfers more closely to Ka. The GO4
expansion parameter $\mathrm{msc}/(q_z^2\mss^2)$, however, reaches $0.30$--$0.37$ at
Ku over $12$--$16\,\mathrm{m\,s^{-1}}$, where a fourth-order truncation becomes
doubtful, whereas the logarithmic form has no truncation parameter.

\begin{table}[t]
  \centering
  \caption{The classical geometric-optics chain and the structure-function
  Kirchhoff model, term by term.}
  \label{tab:concept}
  \footnotesize
  \setlength{\tabcolsep}{3pt}
  \begin{tabular}{p{0.20\columnwidth}p{0.36\columnwidth}p{0.36\columnwidth}}
    \toprule
     & \textbf{GO / GO4} & \textbf{Structure-function Kirchhoff} \\
    \midrule
    State variable & slope PDF and effective curvature &
      $D_n(\bm r)$ at all lags $r$ \\
    Slope variance & GO: filtered $\mss$. GO4: total $\mss$ &
      derived $\msseff=s^2(r^{*})=2\Dtwo(r^{*})/r^{*2}$ \\
    Scale selection & GO: cutoff $k_d=k_r/3$. GO4: effective curvature &
      self-selected $\rstar$ \\
    Radar wavelength & GO: via the chosen cutoff; GO4: via the $\qz$-dependent
      effective-curvature term $\mathrm{msc}/(q_z^2\mss^2)$ (predictive if
      $\mathrm{msc}$ is supplied) &
      via the kernel lag $\rstar$ and the level factor $R_F^2 C_f$ \\
    Profile peakedness & slope kurtosis $c_{40}$ plus GO4 finite-wavelength
      curvature & local non-quadratic $\Dtwo$: log-slope $b$ $+$ optional
      variance-mixture kurtosis $R$ \\
    Band transfer & external spectrum or $\mathrm{msc}(\Uten)$ law; a
      one-band fit does not determine the Ku--Ka split &
      interpolation between anchored kernel lags within the fitted local $\Dtwo$;
      extrapolation outside the anchored interval \\
    Sea state ($\Hs$) & additive $\mss_{\rm sw}$ partition &
      additive tilt $\Delta\sigma^2_{\rm long}(\Uten,\Hs)$ on $\msseff$; hinge from
      the core, size uncalibrated (Sec.~\ref{sec:disc-hs}) \\
    Azimuth & anisotropic, possibly non-Gaussian slope PDF &
      $\Dtwo$ anisotropy ($a_2$) $+$ odd cumulant $D_3$ ($a_1$) \\
    Cost & closed form &
      one $\sim$200-point Hankel quadrature per $(\theta,\Uten)$ \\
    \bottomrule
  \end{tabular}
\end{table}

\subsection{What two frequencies resolve}
\label{sec:disc-ident}
The GO4 comparison raises a more general question: which surface properties can
two near-nadir frequencies constrain? They constrain the level of $s^2(r)$ and, more
weakly, its log-slope $b$. The reason is geometric. Near nadir each band reads $s^2(r)$ around its own kernel lag,
$\rstar_{\rm Ku}\simeq1.9$\,cm and $\rstar_{\rm Ka}\simeq0.7$\,cm
(Fig.~\ref{fig:go4}a). Two points on one curve fix two numbers, its level and its
slope between the points. The level is well determined. The slope $b$ is
determined less well, because the two lags differ only by a factor of about $2.7$
and the logarithmic lever arm between them is short.

A single Ku profile adds little to this. Its incidence dependence does carry
information on $b$, but its sensitivities to the parameters are very unequal. The
Jacobian of the Ku profile in dB, in unscaled $(a,b)$ coordinates with rows weighted by
$\sqrt{n_i}/\langle\sqrt{n_i}\rangle$ over the populated bins in $0$--$11^\circ$ at $4$,
$8$ and $12\,\mathrm{m\,s^{-1}}$, has singular values that differ by a factor of
$40$--$66$. Adding a free level as $\ln L$ raises the ratio to $1200$--$2400$. These
coordinate-dependent ratios show that the Ku profile constrains essentially one
combination of level, $a$ and $b$.

This weak second mode has three consequences. First, band transfer from one band
is unreliable: a Ku-only fit over-predicts the shape difference seven- to ninefold at
$4\,\mathrm{m\,s^{-1}}$: ninefold for the fit of Fig.~\ref{fig:go4}c, started from other node
values, and sevenfold for the fit of Part~II. Second, the retrieved wind
exponent of $b$ moves between $1.45$ and $2.06$ with the objective weights,
against a $95\%$ month-block bootstrap interval of $1.98$--$2.02$ (Part~II). Third,
errors that act unequally on the two bands, such as a residual incidence trend of
the Ka attenuation or a wind-collocation error, are amplified into $b$. Bandwise constant
offsets, in contrast, are absorbed by the levels, while wind-dependent Ku offsets can
still move the fitted nodes.
A third frequency would lengthen the lever arm and sharpen the second mode.

The same limit applies to the kurtosis. Once the band pair fixes $\Dtwo$, the
profile does not require a positive $R$. An incidence-independent quantity, such as
the band reflectivity, can bound $R$, but it carries no information on the profile
shape and so cannot separate $R$ from the scale dependence of $\Dtwo$. Two near-nadir
frequencies therefore do not separate a scale-dependent second cumulant from a
non-Gaussian closure that acts at all lags.

\subsection{Validity envelope across bands}
\label{sec:disc-validity}
Beyond what two frequencies resolve, the transfer to other bands is limited by
both the logarithmic form and the tangent-plane approximation. The roughness
channel enters only through $\rstar$, and the level channel through
$R_{\rm eff}^2$. The kernel lag locates the roughness scales that each band samples. It
does not by itself establish tangent-plane validity, which also requires surface
curvature radii large against the radar wavelength.

The first limit is the logarithmic form itself. Its underlying self-affinity
weakens as wind rises, and by $18\,\mathrm{m\,s^{-1}}$ the local log-slope departs
from a single $b$ by a few percent of $s^2$. Transfer along one $\Dtwo$ curve thus
degrades at the highest winds, even between Ku and Ka.

The second limit is the lag range. W band samples $\rstar\simeq2.4$\,mm and C band
$\rstar\simeq5$\,cm, outside the interval constrained between the Ku and Ka kernel
lags. At W band neither the logarithmic closure nor the tangent-plane approximation
has been tested, and at C band the logarithmic form is extrapolated toward the
decimetre lags at which $s^2(r)$ must flatten. C and W band therefore remain untested
extrapolations. TOPEX recorded C-band backscatter together with Ku
\citep{chapron1995}, as the Jason altimeters have since, but those data are not used here.

\subsection{An explicit sea-state ($\Hs$) term}
\label{sec:disc-hs}
Within this range of validity, sea state can be added as an explicit input, since
altimeters and the SWIM nadir beam measure $\Hs$ over the same or an adjacent
footprint. The long waves tilt the surface, and over the footprint their slopes add
to the short-wave slopes. The baseline $\mss_w$ already contains the long waves of
the reference sea state, so only the departure from it is added. In the GO approximation, the
radar-effective slope variance then reads
\begin{equation}
  \msseff(\Uten,\Hs)=\mss_w(\Uten)+\Delta\sigma^2_{\rm long}(\Uten,\Hs),
  \label{eq:hs-msseff}
\end{equation}
where $\mss_w$ is the effective slope variance at the reference sea state and
$\Delta\sigma^2_{\rm long}$ is the long-wave slope variance in excess of its mean
over the reference window, the same in both bands. This additive form is that of
\citet{tran2007}. Altimeter wind algorithms include $\Hs$ empirically
\citep{gourrion2002}, whereas the published off-nadir and precipitation-radar GMFs
omit it \citep{hossan2021,ren2021}. The anomaly is a slope variance, not a height.
A swell adds height but little slope, so $\Hs$ is only a proxy for
$\Delta\sigma^2_{\rm long}$, which is best taken from the long-wave spectrum.
Long-wave modulation, by comparison, adds only a few-percent correction to the
profile shape through $R=|M|^2\sigma^2_{\rm long}$ (Section~\ref{sec:d4}).

\begin{figure}[t]
  \centering
  \includegraphics[width=0.9\columnwidth]{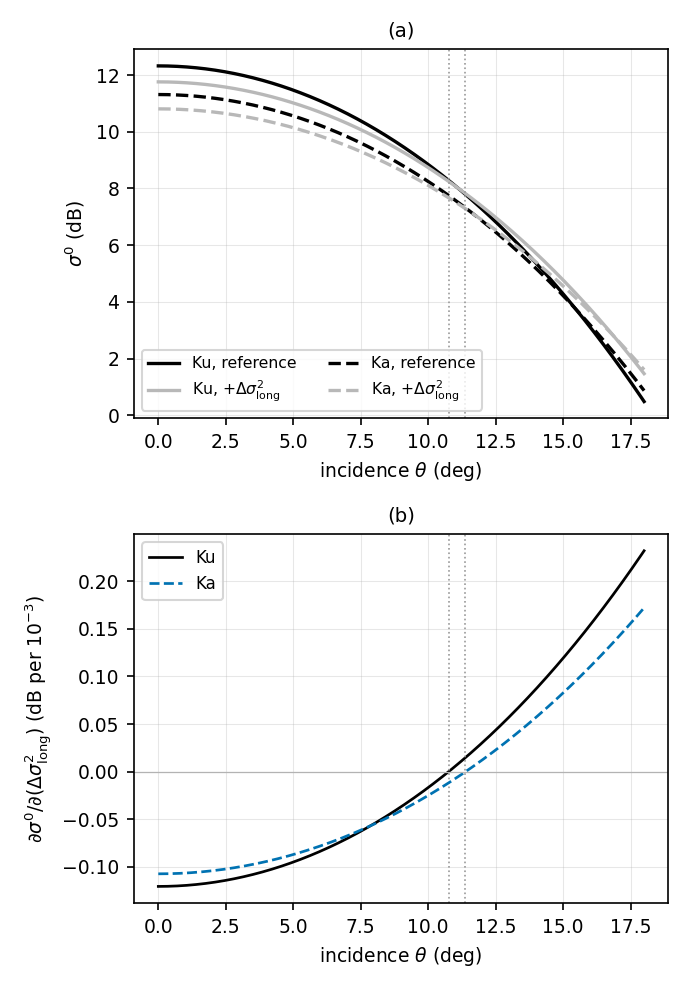}
  \caption{The explicit sea-state term for Ku and Ka from \eqref{eq:hs-msseff} and
  \eqref{eq:hs-deriv} (model only, no fitted coefficient), at $\Uten=10\,\mathrm{m\,s^{-1}}$ with
  Meissner--Wentz reflectivities at $18.3^\circ$C and $\mss_w=s^2(\rstar)$ from the
  joint parameters. (a)~An illustrative band-common anomaly
  $\Delta\sigma^2_{\rm long}=5\times10^{-3}\,\mathrm{rad^2}$ raises $\msseff$ and
  rotates the $\snot(\theta)$ profile near each band's reference-state hinge
  $\tan^2\theta=\mss_w$ (dotted); at nadir it lowers $\snot$ by $0.56\dB$ (Ku) and
  $0.51\dB$ (Ka). (b)~The sensitivity per unit anomaly,
  $\partial\snot_{\rm dB}/\partial(\Delta\sigma^2_{\rm long})
  =(10/\ln10)(\tan^2\theta-\mss_w)/\mss_w^2$, is $-0.120$ (Ku) and
  $-0.107\dB$ (Ka) per $10^{-3}\,\mathrm{rad^2}$ at nadir, and changes sign at
  $10.8^\circ$ (Ku) and $11.4^\circ$ (Ka).}
  \label{fig:hsterm}
\end{figure}

The sea-state sensitivity of the cross-section follows from the geometric-optics
core $\snot\propto\msseff^{-1}\exp(-\tan^2\theta/\msseff)$,
\begin{equation}
  \frac{\partial\snot}{\partial\Hs}
   =\frac{\tan^2\theta-\msseff}{\msseff^{2}}\,\snot\,
    \frac{\partial\Delta\sigma^2_{\rm long}}{\partial\Hs},
  \label{eq:hs-deriv}
\end{equation}
which is the near-nadir form of \citet{tran2007} with
$\partial\Delta\sigma^2_{\rm long}/\partial\Hs>0$. The coefficient
$\partial\Delta\sigma^2_{\rm long}/\partial\Hs$ sets the size of the response, but
not its sign. The sensitivity changes sign at the hinge angle defined by
$\tan^2\theta=\msseff$. A higher sea state lowers the cross-section below that angle
and raises it above (Fig.~\ref{fig:hsterm}). This reproduces the observed reversal of
the $\snot$--$\Hs$ response near $10^\circ$ \citep{tran2007}. To first order in the
anomaly, the sign and the hinge are fixed by the core alone, independently of the
sea-state coefficient; a finite anomaly moves the crossing outward with $\msseff$.

The added slope variance is the same in both bands, but the response differs,
because it is relative to each band's $\msseff$. The larger Ka effective slope
variance reduces the fractional sensitivity and moves the hinge outward, from
$10.8^\circ$ at Ku to $11.4^\circ$ at Ka at $10\,\mathrm{m\,s^{-1}}$ for the joint
parameters. Part~II finds the measured Ku sign change within about $0.5^\circ$ of that of the
full logarithmic Kirchhoff integral, but places the Ka sign change inside the
Ku one, so a band-common tilt does not capture the full
Ka response. At Ku, SWIM supports the qualitative tilt mechanism (Part~II). At fixed wind the response
follows the long-wave slope variance rather than $\Hs$; its amplitude, about half of
\eqref{eq:hs-deriv}, remains unresolved. Its measured size in
$\Hs$ depends on the accuracy of the reference wind, and its calibration as an
operational law is left to a separate study.

\subsection{Research directions}
\label{sec:disc-directions}
The right column of Table~\ref{tab:ledger} lists what the model still takes from
data, and each entry names a tractable problem. The largest concerns the level of
$b(\Uten)$, which the KMC spectrum places $35\%$ (Ku) and $74\%$ (Ka) too high at
$16\,\mathrm{m\,s^{-1}}$. A short-wave source re-tuned over the centimetre band, with
the slope and energy moments preserved, would make $b$ predictive, and stereo
measurements of $D_n(r)$ \citep{mironov2012} would constrain it independently. Referencing
$b$ to the stress-equivalent scatterometer wind changes its nonzero nodes by at most
$4\%$ (Part~II), far less than the KMC level error.

The higher cumulants need a corresponding advance in the modulation model.
Deriving $\lambda_3$ from the MTF phase through \eqref{eq:totalcum}, and $R$ from a
first-principles short-wave relaxation rate, would constrain their long-wave
contributions.

The scattering and surface descriptions can also be extended. A unified small-slope
\citep{voronovich1994} or two-scale kernel would replace the empirical Bragg term
over $11$--$18^\circ$. A globally positive-definite $\Dtwo$ joined to the
logarithmic asymptote would place the closure on a rigorous footing. Finally,
separating a scale-dependent $\Dtwo$ from a non-Gaussian closure would require a
third radar frequency, C band or the $3.2$-GHz S band of Envisat RA-2, which would
also lengthen the lag lever arm; three-frequency C/Ku/Ka altimetry was proposed to
estimate the sea-state bias \citep{amarouche2001}.

\section{Conclusion}
\label{sec:conclusion}

The near-nadir Kirchhoff integral, written on the two-point characteristic function
of the elevation increment, describes ocean backscatter through the structure
functions $D_n(r)$. Its scattering kernel selects the lags it samples, $2.5$--$1.4$\,cm
at Ku and $1.0$--$0.5$\,cm at Ka band over $3$--$20\,\mathrm{m\,s^{-1}}$. Gaussian,
isotropic geometric optics is the quadratic limit of $\Dtwo$, and the filtered
mean-square slope becomes the slope variance $s^2(\rstar)$ at the kernel lag, with
no spectral cutoff.

The central result is one local logarithmic $\Dtwo(r)$ for both bands. Its log-slope
$b(\Uten)$ sets both the incidence shape and the Ku--Ka shape difference, so band
transfer becomes the evaluation of one curve at the second lag. One band does not
fix that transfer: a Ku-only fit over-predicts the shape difference seven- to ninefold
at $4\,\mathrm{m\,s^{-1}}$. Read as a local roughness exponent, $b$ also gives an
effective fractal dimension \eqref{eq:fractaldim} that rises from $2.01$ at $4$ to
$2.10$ at $16\,\mathrm{m\,s^{-1}}$ at the Ku kernel lag.

Physically, $b$ is the level of the short-wave saturation range, and it grows
approximately as the square of the wind speed. Driven by the KMC
spectrum, the integral reproduces that wind exponent, which comes from the base
saturation and not from the breaking terms. The level, however, is $35\%$ (Ku) and
$74\%$ (Ka) too high at $16\,\mathrm{m\,s^{-1}}$. Matched dual-frequency radars thus
constrain the centimetre-scale saturation level that spectral models tune to slope
statistics.

An independently fitted fourth cumulant remains optional. Filtered-slope simulations give a nearly
wind-independent crowding excess kurtosis of $0.13$--$0.18$, but two frequencies resolve
only a level and a weak log-slope mode, and they do not separate a scale-dependent
$\Dtwo$ from a non-Gaussian closure. The sea-state tilt is treated in Section~\ref{sec:disc-hs} and tested in Part~II. Because the roughness is constrained only between
the Ku and Ka kernel lags, transfer to C or W band is extrapolation. The Ku/Ka
realization, KuKa-Mod, and its validation against GPM and CFOSAT/SWIM are developed
in the companion paper.

\section*{Data and Code Availability}
The diagnostic fits use the GPM Dual-frequency Precipitation Radar (DPR) V07A and CFOSAT/SWIM reductions summarized in
the footnote of Section~\ref{sec:bU-origin}, whose full processing is described in
Part~II \citep{part2}. The joint parameter set, an evaluator and look-up tables are released
in the KuKa-Mod v1 record on Zenodo
(\href{https://doi.org/10.5281/zenodo.24084870}{doi:10.5281/zenodo.24084870}); the
binned GPM/SWIM aggregates and the fitting code are not part of that record. The source GPM DPR products
are distributed by NASA/JAXA, and the CFOSAT/SWIM L2 products by CNES/AVISO and CNSA.

\section*{Acknowledgment}
This work was performed as a part of the CFOSAT IFREMER Wind and Wave Operation
Center (IWWOC) development. The IWWOC is co-funded by CNES and IFREMER.

\bibliographystyle{IEEEtranN}
\bibliography{refs}

\end{document}